\documentclass[twocolumn,showpacs,pra,showkeys,superscriptaddress]{revtex4}

\usepackage{graphicx}
\usepackage{dcolumn}
\usepackage{bm}
\usepackage{amsmath,amssymb}
\usepackage{mathbbol,dsfont}
\usepackage{color}

\begin{document}

\title{Degenerate Adiabatic Perturbation Theory: Foundations and Applications}

\author{Gustavo Rigolin}
 \affiliation{Departamento de F\'isica,
Universidade Federal de S\~ao Carlos, S\~ao Carlos, SP 13565-905,
Brazil}
\author{Gerardo Ortiz}
\affiliation{Department of Physics, Indiana University,
Bloomington, IN 47405, USA}
\date{\today}

\begin{abstract}
We present details and expand on the framework
leading to the recently introduced degenerate adiabatic perturbation theory
[Phys. Rev. Lett. \textbf{104}, 170406 (2010)], and on the
formulation of the degenerate adiabatic theorem, along with its
necessary and sufficient conditions given in [Phys. Rev. A \textbf{85}, 062111 (2012)]. 
We start with the adiabatic approximation for degenerate Hamiltonians that paves
the way to a clear and rigorous statement of the associated degenerate
adiabatic theorem, where the non-abelian geometric phase (Wilczek-Zee phase)
plays a central role to its quantitative formulation. We then describe
the degenerate adiabatic perturbation theory, whose
zeroth order term  is the degenerate adiabatic approximation,  in its full generality. 
The 
parameter in the perturbative power series expansion of the time-dependent wave
function is directly associated to the inverse of the time it takes to drive the system
from its initial to its final state. With the aid of the degenerate adiabatic perturbation
theory we obtain rigorous necessary and sufficient conditions for the validity of the
adiabatic theorem of quantum mechanics. Finally, to illustrate the power and 
wide scope of the methodology, we apply the framework
to a degenerate Hamiltonian, whose closed form time-dependent wave function
is derived exactly, and also to other non-exactly-solvable Hamiltonians whose 
solutions are numerically computed. 
\end{abstract}

\maketitle

\section{Introduction}
\label{intro}

It is widely recognized by any practicing theoretical physicist that one seldom
comes across exactly solvable problems in our day to day business.
Most problems have solutions that cannot be expressed
in a simple closed form. Fortunately, there exist at least two strategies one can 
undertake to tackle such problems and still  learn about their solutions. The first one 
is very popular these days due to the increasing processing power of computers.
It is the ``brute force'' approach, the one that tries to solve numerically
the complete problem. This approach has at least one disadvantage though. We cannot
track in a straightforward manner the contributions and relevance 
of the many different interactions among the system's constituents that we are studying. 

The other strategy, when applicable, is generally referred to as ``perturbation theories.'' 
It is particularly
useful when the problem to be solved has a Hamiltonian (or Lagrangian) that can
be split into two parts. One where  the exact solution is known and another 
whose contribution to the overall Hamiltonian is small relative to the first part. 
In this scenario one can express the solution to the original problem 
as a series expansion in terms of the perturbation. 
After achieving a desired accuracy, most of the time dictated by the necessity to
explain experimental data, we may stop the series. If the series is convergent, 
the more terms we keep, the closer is the approximate solution to the exact one. 
All standard perturbation theories, time-dependent or time-independent, 
are more or less akin to this general framework.

Restricting our attention to time-dependent perturbation theories, we notice that they are
built assuming the system's Hamiltonian can be cast as 
$\mathbf{H}(t)=\mathbf{H_0} + \lambda\mathbf{V}(t)$, 
where $\mathbf{H_0}$ is a time-independent term and $\lambda\ll 1$. 
The standard textbook time-dependent perturbation theory \cite{Coh77}, developed by Dirac in
the early days of quantum mechanics, and the Dyson series, are remarkable examples of such
perturbation theories.   

Consider now the following problem. We have a time-dependent Hamiltonian $\mathbf{H}(t)$ that 
\textit{cannot} be written as above, but that varies with time \textit{very slowly}, in 
a sense to be precisely defined later on.  
In this case, 
the evolution of the system is known to be determined by 
the adiabatic approximation (AA) \cite{Mes62,Jan07}. 
In a nutshell, if a system is described by AA, during its whole time evolution,
say $t\in [0,T]$, there are no transitions among different eigenstates 
(for non-degenerate systems) or among different eigenspaces (for degenerate systems).
The time scale $T$ is either an experimental constraint or may be freely chosen,
according to other internal time scales of the system, to make the rate of change 
of $\mathbf{H}(t)$ slow enough. 

To make the statements above quantitative and rigorous, two questions
must be addressed. (a) What are the conditions under which AA is actually a good
approximation to the time evolution of the system? In other words, what is 
the quantitative meaning of slow? (b) If AA is not good enough, what
are its \textit{perturbative} corrections in terms of the rate at which $\mathbf{H}(t)$ is 
changing? From the very beginning one thing is clear though. The standard time-dependent perturbation
theories cannot help us much in addressing rigorously these questions  
since they all assume the time-dependent part of $\mathbf{H}(t)$ is small when compared to 
the time-independent one. Here, we do not assume $\mathbf{H}(t)$ has a  perturbative 
component  or even a time-independent part.

Both questions can be consistently tackled, nevertheless, 
with the aid of adiabatic perturbation theory (APT)
for non-degenerate \cite{Rig08,Pol10a,Pol10b,Pol11a,Pol11b}
and degenerate systems \cite{Rig10}. 
In particular, answers to the first question are what we call
necessary and sufficient conditions for the
validity of the adiabatic theorem (AT) of quantum mechanics.
The necessary condition is correctly handled with the recognition that the geometric
phases, either abelian \cite{Ton10} or non-abelian \cite{Rig12}, are key pieces of
AA. The sufficient conditions as well as perturbative corrections to AA 
in terms of the rate that $\mathbf{H}(t)$ changes are given by 
APT \cite{Rig08} for non-degenerate systems or by the degenerate APT (DAPT) \cite{Rig10} for
degenerate ones.

A main goal of the current article is to present a thorough discussion of DAPT,
highlighting its main differences from standard perturbation theories as well as
its range of validity.
We aim at providing a systematic derivation of all mathematical details that were
omitted in \cite{Rig10}, where DAPT was introduced, and
in \cite{Rig12}, where necessary and sufficient conditions for the validity of the 
degenerate AT (DAT) were presented. In particular, we stress some key properties 
needed to properly develop DAPT that are usually not important in most standard perturbation
theories. We also show that DAPT reduces to APT when no
degeneracy is present and apply all these ideas to a few new examples, where one can
grasp the usefulness of DAPT.

Most importantly, practical quantitative conditions for the validity of the 
adiabatic theorem are also of upmost relevance to the  current problem
of assessing the feasibility of any information processing
scheme that uses the concept of Majorana or Parafermionic 
non-Abelian braiding \cite{Iva01, Cob14}. DAPT is 
in a unique position, as a  theoretical tool,  to estimate potential errors in the physical 
implementation of such topological gates. Furthermore, DAPT can also be employed
to compute analytically the non-adiabatic corrections to the adiabatic population 
transfer and coherent control methods developed in \cite{Una98,Una99,Kis04,Tha04} 
for non-degenerate systems.

To guide the reader to the points she/he is most interested in, 
we divided this article into the following sections.
In Sec. \ref{notation} we present the notation that best suits the mathematical
formulation of DAPT and DAT. Sections \ref{AA} and \ref{DAT} introduces the degenerate 
AA (DAA) and DAT in
their most general form. Section \ref{DAPT} constitutes the core of the manuscript from which
many of the other results follow. It includes the systematic development of DAPT, where
we also discuss its place among other perturbation theories, show its equivalence
to APT when no degeneracy is present, and highlights the key ingredients needed 
to arrive at a consistent perturbation theory.  This  together
with Sec. \ref{notation} are the sections one should read in order to get a general
feeling of DAPT. Section \ref{NSC} gives rigorous as well as practical necessary and sufficient
conditions for the validity of DAT. In Sections \ref{example} and \ref{examples} we work out
several examples that illustrate how DAPT and the conditions for the validity
of DAT should be applied. In Sec. \ref{example}, 
in particular, we present all the calculation details
leading to the exact solution, discovered in [\onlinecite{Rig10}], of a 
time-dependent degenerate Hamiltonian problem 
introduced in [\onlinecite{Bis89}]. Finally, in Sec. \ref{conclusion} we 
summarize the crux of our DAPT approach and briefly  describe our main findings.

\section{A bit of notation}
\label{notation}

In order to keep the equations concise, and 
as similar as possible to the ones for the non-degenerate case \cite{Rig08}, 
we introduce what we call the ``vector of vectors'' or 
``vector of quantum states" notation.
It also helps in simplifying the mathematical manipulations leading to DAPT.
This new object represents all the degenerate states in a single notation. 
For example, a two-fold degenerate ground-eigenspace at a given time $t$ 
has the eigenstates $|0^0(t)\rangle$ and
$|0^1(t)\rangle$. In the ``vector of vectors'' notation one writes
$$
|\mathbf{0}(t)\rangle=
\left(
\begin{array}{c}
|0^0(t)\rangle \\
|0^1(t)\rangle
\end{array}
\right).
$$

In general we have a $d_n$-fold degenerate eigenspace,
$$
|\mathbf{n}(t)\rangle= \left(
\begin{array}{c}
|n^0(t)\rangle \\
|n^1(t)\rangle \\
\vdots \\
|n^{d_n-1}(t)\rangle
\end{array}
\right).
$$
We represent its $g_n$-th element by the following notation,
$$
[|\mathbf{n}(t)\rangle]_{g_n,0} = [|\mathbf{n}(t)\rangle]_{g_n} =
|n^{g_n}(t)\rangle,
$$
where $g_n=0,1,\cdots,d_n-1$.  
In the development of DAPT we often need the transposed vector of quantum
states,
$$
|\mathbf{n}(t)\rangle^{T}= \left(
\begin{array}{cccc}
|n^0(t)\rangle, & |n^1(t)\rangle, & \cdots, & |n^{d_n-1}(t)\rangle
\end{array}
\right).
$$

The ``row vector of bras'' is defined as,
$$\langle\mathbf{
n}(t)|= \left(
\begin{array}{cccc}
\langle n^0(t)|, & \langle  n^1(t)|, & \cdots, & \langle
 n^{d_n-1}(t)|
\end{array}
\right),
$$
and its transpose as
$$
\langle\mathbf{n}(t)|^T= \left(
\begin{array}{c}
\langle n^0(t)| \\
\langle n^1(t)| \\
\vdots \\
\langle n^{d_n-1}(t)|
\end{array}
\right).
$$

In general each eigensubspace $\mathcal{H}_n$ has a 
different number of degenerate states. 
However, it is mathematically convenient to keep the dimension 
fixed ($d_{\sf max}$) and pad zeros whenever the vector belongs to a
 less degenerate eigensubspace. 
In this way, every vector (matrix) that we will be dealing 
with will have the same dimension.
For example, if the ground-eigenspace is
two-fold degenerate and the first excited state three-fold we will have,
$$
|\mathbf{0}(t)\rangle^{T}= \left(
\begin{array}{ccc}
|0^0(t)\rangle, & |0^1(t)\rangle, & 0
\end{array}
\right),
$$
$$
|\mathbf{1}(t)\rangle^{T}= \left(
\begin{array}{ccc}
|1^0(t)\rangle, & |1^1(t)\rangle, & |1^2(t)\rangle
\end{array}
\right).
$$

With such a convention, we can also define a multiplication between 
these objects according to standard matrix multiplication rules. 
Hence, for example, $\langle \mathbf{n}(t)|^T
|\mathbf{m}(t)\rangle^{T}$ is a square matrix. 

We will also find expressions such as
$\mathbf{H}(t) |\mathbf{\Psi}(t)\rangle$. Here it is assumed that the 
Hamiltonian operator $\mathbf{H}(t)$ acts as a ``scalar'' on the vector of 
vectors $|\mathbf{\Psi}(t)\rangle$. In other words, 
$$
\mathbf{H}(t) |\mathbf{\Psi}(t)\rangle = \left(
\begin{array}{c}
\mathbf{H}(t)[|\mathbf{\Psi}(t)\rangle]_0 \\
\mathbf{H}(t)[|\mathbf{\Psi}(t)\rangle]_1 \\
\vdots \\
\mathbf{H}(t)[|\mathbf{\Psi}(t)\rangle]_{d_{\sf max}-1}
\end{array}
\right),
$$
where $d_{\sf max}=\max_{n}\{d_n\}$, is the dimension of the 
most degenerate eigenspace.

Before we move on and to avoid any ambiguity,
it is worth calling attention to two notational issues. 
First, we use the same symbol $T$ to represent transposition of matrices as
well as the total time during which the system's Hamiltonian is evolving. It is easy,
though, to infer which meaning is assigned to it by the context
where it appears. The transposition $T$ is always a superscript while the 
time $T$ is always on the baseline.

Second, many
times throughout this article we will be dealing with the rescaled time
$s=vt$, where $v$ is  the rate of change of the 
Hamiltonian. When formulating DAPT it is convenient to work with
$s$ while when working with the conditions for the validity of DAT it is simpler
to work with $t$. Thus, for instance, $|\mathbf{n}(s)\rangle$,
represents the original vector $|\mathbf{n}(t)\rangle$
with the substitution of  $t$ by $s/v$. Note, however, that the dot always means 
derivative with respect to the argument, i.e.,
$|\dot{\mathbf{n}}(t)\rangle=\partial_t|\mathbf{n}(t)\rangle$ or
$|\dot{\mathbf{n}}(s)\rangle=\partial_s|\mathbf{n}(s)\rangle$. 

\section{Degenerate Adiabatic approximation}
\label{AA}

An unambiguous and quantitative formulation of
DAT must necessary be related to DAA. Briefly, DAT will be shown to be strictly 
connected to the conditions under which DAA is valid. 
In order to present DAA in a clear and consistent manner 
we follow Refs. \cite{Ber84,Ton10,Wil84,Wil11}, where geometric phases 
are at the core of any meaningful AA.

Let us consider an explicitly time-dependent Hamiltonian
$\mathbf{H}(t)$, $t\in[0,T]$, with orthonormal eigenvectors $|n^{g_n}(t)\rangle$.
Each degenerate eigenspace $\mathcal{H}_n$ of 
dimension $d_n$ and eigenenergy $E_n(t)$ possesses $d_n$ degenerate states. Obviously   
\begin{equation}
\mathbf{H}(t)|n^{g_n}(t)\rangle = E_n(t)|n^{g_n}(t)\rangle ,
\label{SE1}
\end{equation}
and we assume that $d_n$ is fixed during the total time 
evolution $T$ (see Fig. \ref{fig1}). An arbitrary initial, $t=0$,  condition
can be written as
\begin{equation}
|\Psi^{(0)}(0)\rangle=\sum_n\sum_{g_n=0}^{d_{n-1}}b_n(0)U_{h_ng_n}^n(0)|n^{g_n}(0)\rangle,
\end{equation}
where $|b_n(0)|^2$ is the probability of finding the system  in eigenspace 
$\mathcal{H}_n$, while $|b_n(0)U_{h_ng_n}^n(0)|^2$ is the probability of measuring a
specific degenerate eigenstate within a given eigenspace. 

The label $h_n$ specifies a particular initial condition
within an eigenspace. If we include all initial conditions 
spanning an orthonormal eigenspace $\mathcal{H}_n$ we arrive at 
the unitary matrix $\mathbf{U}^{n}(0)$, such that 
$\mathbf{U}^{n}(0)(\mathbf{U}^{n}(0))^\dagger = \mathds{1}$.
Then,
an arbitrary state at $t=0$ can be written as
\begin{equation}
\mathbf{|\Psi}^{(0)}(0)\rangle = \sum_{n=0}
b_n(0)\mathbf{U}^{n}(0)\mathbf{|n}(0)\rangle,
\label{BoldPsi0}
\end{equation}
with the usual matrix multiplication rule implied. 
If we want to particularize to a specific initial condition, we just choose the
corresponding element of the vector column $\mathbf{|\Psi}^{(0)}(0)\rangle$.

Using this notation, the most general way of writing 
AA is
\begin{equation}
\mathbf{|\Psi}^{(0)}(t)\rangle = \sum_{n=0}
\mathrm{e}^{-\mathrm{i}\omega_n(t)}
b_n(0)\mathbf{U}^{n}(t)\mathbf{|n}(t)\rangle, \label{vector0}
\end{equation}
with dynamical phase
\begin{equation}
\omega_n(t)= \frac{1}{\hbar}\int_{0}^{t}E_n(t')dt' ,
\label{dynamical}
\end{equation}
and unitary matrix
$\mathbf{U}^{n}(t)$
given by the non-abelian
Wilczek-Zee phase (WZ phase) \cite{Wil84},
\begin{equation}
\mathbf{U}^{n}(t) = \mathbf{U}^{n}(0)\mathcal{T}
\exp\left( \int_0^t\mathbf{A}^{nn}(t')dt'\right),
\label{WZphase}
\end{equation}
where $\mathcal{T}$ denotes a time-ordered exponential,
$\mathbf{A}^{mn}(t)=-\mathbf{M}^{mn}(t)$,
and
\begin{equation}
[\mathbf{M}^{mn}(t)]_{g_mh_n}=M^{nm}_{h_ng_m}(t)=\langle n^{h_n}(t)| \dot{m}^{g_m}(t)\rangle,
\label{M}
\end{equation}
a $d_m \times d_n$ matrix. Note how the subscripts and superscripts are defined from one 
equality to the other in Eq. (\ref{M}). With the vector of vectors notation 
$$[\mathbf{M}^{mn}(t)]^T =[\langle \mathbf{n}(t)|^T|\dot{\mathbf{m}}(t)\rangle^{T}].$$

Whenever $d_n=1$, the 
eigenspace $\mathcal{H}_n$ has no degeneracy and Eq. (\ref{WZphase})
reduces to the exponential of the abelian Berry phase. Thus, Eq. (\ref{vector0}) is the most
general way of writing AA for degenerate as well as non-degenerate systems.
The physical meaning of AA is clear, the system evolves without transitions between
eigenspaces but within each eigenspace the relative weights of each degenerate
eigenstate is dictated by the WZ phase. 

Had we started at a particular eigenstate, say the ground state
$|0^0(0)\rangle$ for definiteness, we would have 
\begin{equation}
b_n(0) = \delta_{n0} \hspace{1cm} \mbox{and} \hspace{1cm}
\mathbf{U}^{n}(0) = \mathds{1}. \label{initialvector}
\end{equation}
For $b_n(0)$ and $\mathbf{U}^{n}(0)$
given above, the first element of $\mathbf{|\Psi}^{(0)}(t)\rangle$ will
give the time evolution for the initial condition $|0^0(0)\rangle$, the
second for $|0^1(0)\rangle$ and so on. Using
Eq.~(\ref{initialvector}) we have for DAA
\begin{equation}
\mathbf{|\Psi}^{(0)}(t)\rangle =
\mathrm{e}^{-\mathrm{i}\omega_0(t)}
\mathbf{U}^{0}(t)\mathbf{|0}(t)\rangle, 
\label{vector00}
\end{equation}
which implies
\begin{displaymath}
[\mathbf{|\Psi}^{(0)}(t)\rangle]_{0}=|\Psi^{(0)}(t)\rangle =
\sum_{g_0=0}^{d_0-1} \mathrm{e}^{-\mathrm{i}\omega_{0}(t)}
U^0_{0g_0}(t)|0^{g_0}(t)\rangle,
\end{displaymath}
if we focus on the first element of the vector
$\mathbf{|\Psi}^{(0)}(t)\rangle$, i.e., the initial
condition $|0^0(0)\rangle$ evolved up to time $t$.
The reader is directed to appendix \ref{appendixA} 
for a derivation of DAA,  
with the WZ phase naturally appearing.

The rigorous version of DAT presented in Sec. \ref{DAT}
is nothing but a statement about the validity of DAA as given by Eq.~(\ref{vector0}),
in such a way that necessary and sufficient conditions to its validity can be formulated and
proved. Moreover, it is presented in a way such that when the system's dynamics cannot
be approximated by DAA, DAPT of Sec. \ref{DAPT} furnishes perturbative corrections 
 in a consistent fashion. 

\section{Degenerate adiabatic theorem}
\label{DAT}

The dynamics of a closed quantum system is generally governed by the Schr\"odinger 
equation (SE) 
\begin{equation}
\mathrm{i}\,\hbar\, |\dot{\mathbf{\Psi}}(t)\rangle =
\mathbf{H}(t) |\mathbf{\Psi}(t)\rangle.
\label{SE}
\end{equation}
The 
DAT  sets the conditions under which DAA
holds, or equivalently, sets the conditions on the rate of change of the Hamiltonian that
makes DAA a good approximation to the solution of the SE.
In its most general formulation DAT can be presented as follows.

\begin{quote}
If a system's Hamiltonian $\mathbf{H}(t)$ changes \textit{slowly} 
during the course of
time, say from $t=0$ to $t=T$, and the system is prepared in an arbitrary superposition
of eigenstates of $\mathbf{H}(t)$ at $t=0$ (Eq.~(\ref{BoldPsi0})),
then the transitions between eigenspaces
$\mathcal{H}_n$ of $\mathbf{H}(t)$ during the interval $t\in [0,T]$ are \textit{negligible} 
and the system \textit{evolves} according to DAA (Eq.~(\ref{vector0})).
\end{quote}

\begin{figure}[!ht]
\includegraphics[width=8cm]{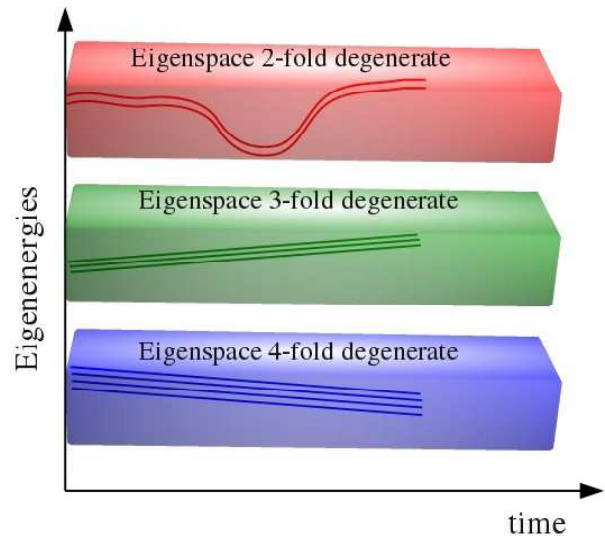}
\caption{\label{fig1} (Color online) Example of spectral time evolution. 
For DAT and DAPT be meaningful, each eigenspace
$\mathcal{H}_n$ can have an arbitrary number of \textit{fixed} degeneracies, 
but there must always be a gap, constant or not, between different eigenspaces. 
In DAA, no transitions occur between eigenspaces while 
within each eigenspace the system acquires the  non-abelian WZ
phase.}
\end{figure}

Note that this statement also applies for non-degenerate systems. DAT as given
above 
is more general than we usually see in the literature since we allow
the system to start at an arbitrary superposition of eigenspaces. 
After the system starts evolving DAT only tells us that no transition
occurs between states with different energies. Within a given eigenspace,
these transitions are given by the WZ-phase. See Fig. \ref{fig1} for an
schematic representation of DAT.

For the sake of comparison, we state DAT in a form that resembles the standard
way of presenting AT for non-degenerate systems, namely, when one starts at 
a given eigenstate of the system (or in a given eigenspace in the degenerate
scenario). Assuming, without loss of generality, that the system begins at
the ground eigenspace $|\mathbf{0}(0)\rangle$, DAT reads:

\begin{quote}
If a system's Hamiltonian $\mathbf{H}(t)$ changes \textit{slowly} 
during the course of
time, say from $t=0$ to $t=T$, and the system is prepared at the 
ground eigenspace of $\mathbf{H}(t)$ at $t=0$,
then the system \textit{evolves} according to $|\mathbf{\Psi}^{(0)}(t)\rangle =
\mathrm{e}^{-\mathrm{i}\omega_0(t)}
\mathcal{T}
\exp\left( \int_0^t\mathbf{A}^{nn}(t')dt'\right)|\mathbf{0}(t)\rangle.$
\end{quote}

 To avoid any possible misunderstanding it is worth stressing the following: 
DAA is based on the assumption that the 
rate of change of $\mathbf{H}(t)$ is {\it slow}. Intuitively, the 
latter notion can be understood as a relation between a characteristic {\it internal} 
time  $T_i$ of the evolved system (the inverse of 
its characteristic frequency, for instance) encoded in $\mathbf{H}(t)$  
and the total evolution time $T$, related to the time it takes to
drive $\mathbf{H}(t)$ throughout the parameter space to its final
destination. It is the interplay between these
internal and external times that dictates whether DAA is a good approximation for
the system time evolution. For a given path in parameter space  
one can always tune $\mathbf{H}(t)$ to make the system adiabatic, although for some Hamiltonians 
this will reflect in an evolution time $T$ prohibitively large in a
possible experimental implementation. 
In other words, one can always decrease the rate of change of $\mathbf{H}(t)$
by increasing the time to drive $\mathbf{H}(t)$ to its final configuration in parameter space.
This state of affairs, although intuitive, is not satisfactory 
from a mathematical standpoint since it provides no 
quantitative notion of \textit{slowness}. This lack of precise
meaning of slow  is a main 
source of controversy. By using DAPT \cite{Rig10}, a generalization of
APT \cite{Rig08}, we can give a precise meaning to this notion of 
slowness, which is crucial for the derivation of the necessary and 
sufficient conditions for the validity of DAT. 

Therefore, before we state and prove the necessary and sufficient conditions for the validity  of 
DAT, we need
first to present DAPT in its full generality and details. DAPT is the tool we need 
to continue the discussion about DAT and also the correct way to obtain higher order corrections
to DAA. Moreover, as we will see, the geometric WZ phase and DAA 
will appear naturally as the zeroth order term of DAPT.

\section{Degenerate adiabatic perturbation theory}
\label{DAPT}

\subsection{The ansatz}

An important characteristic of DAPT is its practical utility. 
As we will see, DAPT can be actually employed to systematically 
approximate any time-dependent 
problem whose snapshot Hamiltonian can be efficiently diagonalized.
Also, DAPT is, to the best of our knowledge, the first perturbation theory 
specially designed for \textit{degenerate} systems about  AA. 
Mathematically, it is a series expansion in terms of the small adiabatic parameter $$v=1/T,$$ 
representing the rate of change of $\mathbf{H}(t)$. Moreover, higher order
terms of the perturbative series are recursively obtained from  their 
lower order terms.  

As extensively discussed for the non-degenerate case \cite{Rig08}, the
usefulness and success of APT are primarily connected to the choice of the {\it
right ansatz} for the form of the solution to SE.  This can be seen by noticing
that the perfect ansatz would {\it factor out} the dependence of
all terms of order $\mathcal{O}(v^{0})$ and below.
In particular, terms $\mathcal{O}(v^{-1})$ and below are the problematic ones
when $v\rightarrow 0$ and should be handled with caution.  In addition to
this, an important insight behind the degenerate ansatz is the recognition
that we have non-abelian phases, 
which are represented by matrices, and a dynamical phase.  
We have somehow to explore all these facts at the very
beginning of the construction of the ``vector of vectors'' ansatz of DAPT in order
to make it work successfully. 

Let us write down the ansatz and then explain the
quantities appearing in it. We assume that the solution to the
time-dependent SE can be written as
\begin{equation}
|\mathbf{\Psi}(s)\rangle =
\sum_{p=0}^{\infty}v^p|\mathbf{\Psi}^{(p)}(s)\rangle,
\label{ansatz}
\end{equation}
where
\begin{equation}
|\mathbf{\Psi}^{(p)}(s)\rangle = \sum_{n=0}
\mathrm{e}^{-\frac{\mathrm{i}}{v}\omega_n(s)}
\mathbf{B}_{n}^{(p)}(s)|\mathbf{n}(s)\rangle 
\label{ansatzA}
\end{equation}
and
\begin{equation}
\mathbf{B}_n^{(p)}(s) = \sum_{m=0}
\mathrm{e}^{\frac{\mathrm{i}}{v}\omega_{nm}(s)}
\mathbf{B}_{mn}^{(p)}(s). \label{ansatzB}
\end{equation}
Here $\omega_{nm}(s)=\omega_{n}(s)-\omega_{m}(s)$,
$|\mathbf{\Psi}(s)\rangle$ and
$|\mathbf{\Psi}^{(p)}(s)\rangle$ are column vectors of dimension
$d_{\sf max}=\max_{n}\{d_n\}$, while
$\mathbf{B}_{n}^{(p)}(s)$ and $\mathbf{B}_{mn}^{(p)}(s)$ are
matrices with dimensions $d_n \times d_n$ and $d_m \times d_n$,
with $d_n$ and $d_m$ being the level of degeneracy of eigenspaces
$\mathcal{H}_n$ and $\mathcal{H}_m$. 
We call each element of the vectors and matrices as
$[\mathbf{|\Psi}(s)\rangle]_{g_n,0} =
[\mathbf{|\Psi}(s)\rangle]_{g_n}
$,
$
[\mathbf{B}_{mn}^{(p)}(s)]_{h_mg_n}
$
and
$
[\mathbf{B}_{n}^{(p)}(s)]_{h_ng_n}.
$

Note that we are now working with the rescaled time $s=vt$, $s\in [0,1]$ for
$t\in [0,T]$. This 
is crucial to correctly identify the order $v$ of each term appearing in
the perturbative series expansion. In the rescaled time the dynamical phase
is 
\begin{equation}
\omega_n(s)= \frac{1}{\hbar}\int_{0}^{s}E_n(s')ds'.
\label{dynamicalS}
\end{equation}
As we said in Sec. \ref{notation}, when we write $E_n(s)$
we mean that the function $E_n(t)$, or any other function of $t$,
is written changing every $t$ by $s/v$. 
Hence, for example, $E_n(t)=at^3$ gives $E_n(s)=as^3/v^3$.

Equation (\ref{ansatz}) tells us that the solution to SE is 
expressed as a series expansion in the perturbative parameter $v$,
with each order given by Eq.~(\ref{ansatzA}). Had we stopped at
Eq.~(\ref{ansatzA}) we would have arrived at a deadlock after inserting
the ansatz into SE. In order to make progress and get a recursive relation
that gives order $p+1$ coefficient as function of order $p$, Eq.~(\ref{ansatzB})
is of utmost importance. Putting all these pieces together the ansatz can be 
written as
\begin{equation}
\mathbf{|\Psi}(s)\rangle = \sum_{n,m=0}\sum_{p=0}^{\infty} v^p
\mathrm{e}^{-\frac{\mathrm{i}}{v}\omega_{m}(s)}
\mathbf{B}_{mn}^{(p)}(s)\mathbf{|n}(s)\rangle. \label{ansatz2}
\end{equation}
                                                                   
Before we proceed, it is important to explain the physical meaning of the small
parameter $v=1/T$, whose choice is related to the ``velocity'' or rate at which
$\mathbf{H}(s)$ changes with time. DAPT is best suited for a system in which 
$\mathbf{H}(s)=\mathbf{H}(\mathbf{r}(s))$ goes from an initial configuration at $s=0$ to a final
one at $s=t/T=1$. Its change is driven by the evolution of $\mathbf{r}(s)$
throughout the parameter space, which can
be, for instance, a varying external field or an internal coupling constant. 
The choice of $v$, or equivalently the total time of the experiment $T$, making 
DAPT convergent depends on how fast or slow $\mathbf{r}(s)$ changes with time. 
It may happen that DAPT does not converge for a particular combination of the values of
the changing rate for $\mathbf{r}(s)$ and 
the total time $T$ employed to drive the system to the 
desired configuration in the parameter space. This fact simply implies 
that AA is not a good approximation for the system's whole evolution.
However, by properly slowing down how $\mathbf{r}(s)$ evolves to its final desired 
configuration, which subsequently increases the duration $T$ of the experiment, 
one can make DAPT converge and guarantee 
AA to be a good description to the system's \textit{whole} evolution from $t=0$ to $t=T$.
More details about the meaning of $v$ are given when we apply DAPT to
several examples in Secs. \ref{example} and \ref{examples}.           

\subsection{Initial conditions}

In addition to the ansatz that correctly highlights each order $v$ and
factors out the dynamical phase, where terms $\mathcal{O}(1/v)$ and below are present,
DAPT can only be realized if the initial conditions
of the system are taken into account. This is a characteristic of DAPT that
differentiates it from all standard perturbation theories, where initial conditions
only play a secondary role. Here, however, initial conditions play a central
role. When properly handled it 
introduces additional terms to the perturbation series, without which DAPT fails. 

The initial conditions and constraints 
imposed on the ansatz must  satisfy:
\begin{itemize}
\item[(i)] The zeroth order of DAPT ($p=0$) must be such that no transitions
between different eigenspaces occur.
\item[(ii)] For $p\geq 1$ we must have
$\mathbf{|\Psi}^{(p)}(0)\rangle = 0$.
\end{itemize}

Condition (i) implies 
\begin{equation}
\mathbf{B}_n^{(0)}(s) = b_n(0)\mathbf{U}^{n}(s) \Longrightarrow
\mathbf{B}_{mn}^{(0)}(s) = b_n(0)\mathbf{U}^{n}(s)\delta_{nm},
\label{B0}
\end{equation}
which leads to (cf. Eq.~(\ref{vector0}))
\begin{equation}
\mathbf{|\Psi}^{(0)}(s)\rangle = \sum_{n=0}
\mathrm{e}^{-\frac{\mathrm{i}}{v}\omega_n(s)}
b_n(0)\mathbf{U}^{n}(s)\mathbf{|n}(s)\rangle, \label{vector0S}
\end{equation}
where we assume nothing about
$\mathbf{U}^n(s)$ but the fact that it is a unitary matrix
at $s=0$. Also, at $s=0$ we recover the initial wave function
(\ref{BoldPsi0}).
Condition (ii) in turn implies for $p\geq 1$,
\begin{equation}
\mathbf{B}_n^{(p)}(0) = 0 \Longrightarrow \mathbf{B}_{nn}^{(p)}(0)
= - \mathop{\sum_{m=0}}_{m\neq n} \mathbf{B}_{mn}^{(p)}(0).
\label{Bnn0}
\end{equation}

\subsection{The recursive equations}

In order to get recursive relations for the matrices 
$\mathbf{B}_{mn}^{(p)}(s)$ we must work with the
transposed ansatz,
\begin{equation}
\mathbf{|\Psi}(s)\rangle^T = \sum_{n,m=0}\sum_{p=0}^{\infty} v^p
\mathrm{e}^{-\frac{\mathrm{i}}{v}\omega_{m}(s)}
\mathbf{|n}(s)\rangle^T\mathbf{B}_{mn}^{(p)}(s)^T
\label{ansatzT}
\end{equation}
and the transposed SE
\begin{equation}
\mathrm{i}\hbar v \frac{\mathrm{d}}{\mathrm{d}s}
\mathbf{|\Psi}(s)\rangle^T = \mathbf{H}(s)
\mathbf{|\Psi}(s)\rangle^T. \label{SET}
\end{equation}
Inserting Eq.~(\ref{ansatzT}) into
Eq.~(\ref{SET}) we get
\begin{eqnarray}
\sum_{n,m=0}\sum_{p=0}^{\infty} & 
v^p &\mathrm{e}^{-\frac{\mathrm{i}}{v}\omega_{m}(s)}
\left(\frac{\mathrm{i}}{\hbar
v}\Delta_{nm}(s)\mathbf{|n}(s)\rangle^T\mathbf{B}_{mn}^{(p)}(s)^T
\right. \nonumber \\
&+&\left. |\dot{\mathbf{n}}(s)\rangle^T\mathbf{B}_{mn}^{(p)}(s)^T
+\mathbf{|n}(s)\rangle^T\dot{\mathbf{B}}_{mn}^{(p)}(s)^T
\right)=0, \nonumber \\
\label{vector1}
\end{eqnarray}
where
\begin{equation}
\Delta_{nm}(s)=E_n(s) - E_m(s).
\end{equation}

Now, if we left multiply Eq.~(\ref{vector1}) by the column vector of bras
$\langle\mathbf{k}(s)|^T$ we get after using 
$\langle\mathbf{k}(s)|^T |\mathbf{n}(s)\rangle^T=\delta_{kn}\mathds{1}$,
\begin{eqnarray}
\sum_{m=0}\sum_{p=0}^{\infty} &
v^p &
\mathrm{e}^{-\frac{\mathrm{i}}{v}\omega_{m}(s)}
\left(\frac{\mathrm{i}}{\hbar
v}\Delta_{km}(s)\mathbf{B}_{mk}^{(p)}(s)^T
+\dot{\mathbf{B}}_{mk}^{(p)}(s)^T\right. \nonumber \\
&+&\left. \sum_{n=0}\mathbf{M}^{nk}(s)^T\mathbf{B}_{mn}^{(p)}(s)^T
\right)=0. \label{vector2}
\end{eqnarray}
The first term $\sum_{m}\sum_{p}v^p
\mathrm{e}^{-\frac{\mathrm{i}}{v}\omega_{m}(s)} \mathrm{i}\Delta_{km}(s)\mathbf{B}_{mk}^{(p)}(s)^T/\hbar v$
in Eq.~(\ref{vector2}) can be cast as follows if we explicitly isolate the $p=0$ term,
\begin{eqnarray}
\sum_{m=0}\sum_{p=0}^{\infty} & v^p
& \mathrm{e}^{-\frac{\mathrm{i}}{v}\omega_{m}(s)}
\frac{\mathrm{i}}{\hbar}\Delta_{km}(s)\mathbf{B}_{mk}^{(p+1)}(s)^T
\nonumber \\
&+& \sum_{m=0} \mathrm{e}^{-\frac{\mathrm{i}}{v}\omega_{m}(s)}
\frac{\mathrm{i}}{\hbar
v}\Delta_{km}(s)\mathbf{B}_{mk}^{(0)}(s)^T.
\label{divterm}
\end{eqnarray}

Employing the initial condition, Eq.~(\ref{B0}), we readily see
that the last sum in (\ref{divterm}) is zero and SE is satisfied if the 
remaining terms multiplying  $v^p \mathrm{e}^{-\frac{\mathrm{i}}{v}\omega_{m}(s)}$ 
in (\ref{vector2}) are zero.
This leads to the following recursive condition,
\begin{eqnarray}
\frac{\mathrm{i}}{\hbar}\Delta_{nm}(s)\mathbf{B}_{mn}^{(p+1)}(s)
+\dot{\mathbf{B}}_{mn}^{(p)}(s)
+\sum_{k=0}\mathbf{B}_{mk}^{(p)}(s)\mathbf{M}^{kn}(s)
= 0, \nonumber \\
\label{recursiveB}
\end{eqnarray}
where we have swapped the indexes $k \leftrightarrow n$ and
taken the transpose.
This is the main recursive relation, from which we are able to
compute $\mathbf{B}_{mn}^{(p)}(s)$ to all orders in $v=1/T$ and give consistent
successive corrections to DAA. As we show in Appendix
\ref{appendixB}, it reduces to the recursive relation obtained in \cite{Rig08} 
for non-degenerate Hamiltonians.

It is worth noting that the initial condition (\ref{B0}) was crucial
to cancel the second term of (\ref{divterm}), whose limit as $v\rightarrow 0$
diverges. This highlights  the importance of the initial conditions on the
development of DAPT. In what follows, we will encounter another instance where
the other piece of the initial conditions, Eq.~(\ref{Bnn0}), becomes relevant.

\subsection{The zeroth and first order coefficients}

We now explicitly compute $\mathbf{B}_{mn}^{(p)}(s)$ up to
first order, i.e., we need to consider the instances where $p=0$ and $p=1$
with either $m=n$ or $m\neq n$. 

\subsubsection{$p=0$ and $m=n$}

In this case  Eq.~(\ref{recursiveB}) becomes,
\begin{eqnarray}
\dot{\mathbf{B}}_{nn}^{(0)}(s)
+\sum_{k=0}\mathbf{B}_{nk}^{(0)}(s)\mathbf{M}^{kn}(s)
= 0,
\end{eqnarray}
and using Eq.~(\ref{B0}) we get
\begin{eqnarray}
b_n(0)\left( \dot{\mathbf{U}}^n(s) +
\mathbf{U}^n(s)\mathbf{M}^{nn}(s)
\right) = 0. \label{p0nequalsm}
\end{eqnarray}
Since, in general, $b_n(0)\neq 0$
the term inside the parenthesis must necessarily be zero and it becomes
$\dot{\mathbf{U}}^n(s) =
\mathbf{U}^n(s)\mathbf{A}^{nn}(s)$, where $\mathbf{M}^{nn}(s)=
-\mathbf{A}^{nn}(s)$. The formal solution to that 
equation is exactly the WZ-phase, Eq.~(\ref{WZphase}). 
In other words,  
the WZ-phase naturally appears in the development of DAPT, 
as anticipated in previous sections.
It is the solution to the zeroth order recursive equation
supplemented with the correct initial condition.

\subsubsection{$p=0$ and $m\neq n$}

Now Eq.~(\ref{recursiveB}) becomes
\begin{eqnarray}
\frac{\mathrm{i}}{\hbar}\Delta_{nm}(s)\mathbf{B}_{mn}^{(1)}(s)
+\sum_{k=0}\mathbf{B}_{mk}^{(0)}(s)\mathbf{M}^{kn}(s)
= 0,
\end{eqnarray}
and using Eq.~(\ref{B0}) we get
\begin{equation}
\mathbf{B}^{(1)}_{mn}(s) =
\frac{\mathrm{i}\hbar}{\Delta_{nm}(s)}\mathbf{U}^m(s)\mathbf{M}^{mn}(s)b_m(0),
\hspace{.25cm} m \neq n. \label{p0ndifferentm}
\end{equation}

\subsubsection{$p=1$ and $m=n$}

In this scenario we can write Eq.~(\ref{recursiveB}) as follows,
\begin{eqnarray*}
\dot{\mathbf{B}}_{nn}^{(1)}(s)
+\mathbf{B}_{nn}^{(1)}(s)\mathbf{M}^{nn}(s)
+\mathop{\sum_{k=0}}_{k \neq
n}\mathbf{B}_{nk}^{(1)}(s)\mathbf{M}^{kn}(s)
= 0.
\end{eqnarray*}
To solve this equation we make the following change of
variables,
\begin{equation}
\mathbf{B}_{nn}^{(1)}(s) =
\mathbf{\tilde{B}}_{nn}^{(1)}(s)\mathbf{U}^n(s),
\label{change}
\end{equation}
which leads to
\begin{eqnarray*}
\mathbf{\tilde{B}}_{nn}^{(1)}(s)\left( \dot{\mathbf{U}}^n(s)  
+ \mathbf{U}^n(s)\mathbf{M}^{nn}(s)
\right)
+
\mathbf{\dot{\tilde{B}}}_{nn}^{(1)}(s)\mathbf{U}^n(s)
\nonumber \\
+ \mathop{\sum_{k=0}}_{k \neq n}
\mathbf{B}_{nk}^{(1)}(s)\mathbf{M}^{kn}(s) = 0.
\end{eqnarray*}
The term inside the parenthesis is zero (WZ-phase), and 
using the unitarity of $\mathbf{U}^n(s)$ we can solve
for $\mathbf{\tilde{B}}_{nn}^{(1)}(s)$,
\begin{equation}
\mathbf{\tilde{B}}_{nn}^{(1)}(s) =
\mathbf{\tilde{B}}_{nn}^{(1)}(0) -\!
\mathop{\sum_{m=0}}_{m \neq n}\!\int_0^s
\!\!\!\mathbf{B}_{nm}^{(1)}(s')
\mathbf{M}^{mn}(s')
\!\left(\mathbf{U}^n(s')\right)^\dagger
\!\!\mathrm{d}s', \label{almost_p1nequalsm}
\end{equation}
where we have changed $k \rightarrow m$.
Next we need to express the initial condition (\ref{Bnn0}) in
terms of the new variable $\mathbf{\tilde{B}}_{nn}^{(1)}(s)$. 
Using the unitarity of $\mathbf{U}^n(s)$ and
Eq.~(\ref{p0ndifferentm}) we can write Eq.~(\ref{Bnn0}) for $p=1$
as
\begin{equation}
\mathbf{\tilde{B}}_{nn}^{(1)}(0) = -\mathrm{i}\hbar
\mathop{\sum_{m=0}}_{m \neq n}
\frac{
\mathbf{U}^m(0)
\mathbf{M}^{mn}(0)
\left(\mathbf{U}^n(0)\right)^\dagger
}
{\Delta_{nm}(0)}
b_m(0). \label{initialBtilde}
\end{equation}
Finally, inserting Eqs.~(\ref{p0ndifferentm}) and
(\ref{initialBtilde}) into (\ref{almost_p1nequalsm}) and the
result into (\ref{change}) we  get
\begin{eqnarray}
\mathbf{B}_{nn}^{(1)}(s) &=& -\mathrm{i}\hbar
\mathop{\sum_{m=0}}_{m \neq n}
\frac{\mathbf{U}^m(0)
\mathbf{M}^{mn}(0)
\left(\mathbf{U}^n(0)\right)^\dagger
\mathbf{U}^n(s)}
{\Delta_{nm}(0)}
b_m(0) 
\nonumber \\
&&+\mathrm{i}\hbar \mathop{\sum_{m=0}}_{m \neq n}
\mathbf{J}^{nmn}(s)\mathbf{U}^n(s) b_{n}(0),
\label{p1nequalsm}
\end{eqnarray}	
where we define 
\begin{equation}
\mathbf{J}^{nmn}(s) =\int_0^s\!\!\mathrm{d}s'\!\!
\left(\frac{\mathbf{U}^n(s')\mathbf{M}^{nm}(s')\mathbf{M}^{mn}(s')
\left(\mathbf{U}^n(s')\right)^\dagger}{\Delta_{nm}(s')}\right)\!.
\label{J}
\end{equation}

Note that the initial condition (\ref{Bnn0}) is responsible for 
the first term of Eq.~(\ref{p1nequalsm}) and the second term
depends on the history (integration over time) of the evolution
of the system. In the Appendix \ref{appendixC} we show how to
obtain the general solution of Eq.~(\ref{recursiveB}), i.e.,
we provide an explicit expression for $\mathbf{B}_{nm}^{(p+1)}(s)$ in terms
of $\mathbf{B}_{nm}^{(p)}(s)$.  

\subsection{The zeroth and first order corrections}

With the previous coefficients we are able to write down 
the zeroth and first order wave functions that approximate the
exact solution to SE according to DAPT.

\subsubsection{The zeroth order term}

The zeroth order term in the expansion is DAA since we have shown that
DAPT implies that $\mathbf{U}^{n}(s)$ is the WZ-phase.  
For $p=0$ if we insert Eq.~(\ref{B0}) into (\ref{ansatz2}) 
we get
\begin{equation}
\mathbf{|\Psi}^{(0)}(s)\rangle=\sum_{n=0}
\mathrm{e}^{-\frac{\mathrm{i}}{v}\omega_{n}(s)}
b_n(0)\mathbf{U}^{n}(s)\mathbf{|n}(s)\rangle.
\end{equation}

Starting at the ground state $|0^0(0)\rangle$ we must add
condition (\ref{initialvector}) and we obtain
\begin{equation}
\mathbf{|\Psi}^{(0)}(s)\rangle=
\mathrm{e}^{-\frac{\mathrm{i}}{v}\omega_{0}(s)}
\mathbf{U}^{0}(s)\mathbf{|0}(s)\rangle.
\end{equation}
The first element of the vector above expresses 
the zeroth order term for the initial condition $|0^0(0)\rangle$ as,
\begin{eqnarray}
[\mathbf{|\Psi}^{(0)}(s)\rangle]_0 &=&
|\Psi^{(0)}(s)\rangle \nonumber \\
&=&\mathrm{e}^{-\frac{\mathrm{i}}{v}\omega_{0}(s)}
[\mathbf{U}^{0}(s)\mathbf{|0}(s)\rangle]_0 \nonumber \\
&=&\mathrm{e}^{-\frac{\mathrm{i}}{v}\omega_{0}(s)}
\sum_{g_0=0}[\mathbf{U}^{0}(s)]_{0g_0}[\mathbf{|0}(s)\rangle]_{g_00}
\nonumber \\
&=& \mathrm{e}^{-\frac{\mathrm{i}}{v}\omega_{0}(s)}
\sum_{g_0=0}U^{0}_{0g_0}(s)|0^{g_0}(s)\rangle. \label{zeroth}
\end{eqnarray}

\subsubsection{The first order correction}

Setting $p=1$ in Eq.~(\ref{ansatzA}) we can write it as
\begin{eqnarray}
\mathbf{|\Psi}^{(1)}(s)\rangle&=&\sum_{n=0}
\mathrm{e}^{-\frac{\mathrm{i}}{v}\omega_{n}(s)}
\mathbf{B}_{nn}^{(1)}(s)\mathbf{|n}(s)\rangle
\nonumber \\
&&+ \mathop{\sum_{n,m=0}}_{m\neq n}
\mathrm{e}^{-\frac{\mathrm{i}}{v}\omega_{m}(s)}
\mathbf{B}_{mn}^{(1)}(s) \mathbf{|n}(s)\rangle.
\end{eqnarray}
Inserting Eqs.~(\ref{p0ndifferentm}) and (\ref{p1nequalsm}) 
we get
\begin{widetext}
\begin{eqnarray}
|\mathbf{\Psi}^{(1)}(s)\rangle&=&
\mathrm{i}\hbar \mathop{\sum_{n,m=0}}_{m \neq
n}\mathrm{e}^{-\frac{\mathrm{i}}{v}\omega_n(s)}b_{n}(0)
\mathbf{J}^{nmn}(s) \mathbf{U}^n(s)\mathbf{|n}(s)\rangle
-\mathrm{i}\hbar \mathop{\sum_{n,m=0}}_{m \neq n}
\mathrm{e}^{-\frac{\mathrm{i}}{v}\omega_n(s)}b_{m}(0)
\frac{\mathbf{U}^m(0)\mathbf{M}^{mn}(0)
\left(\mathbf{U}^n(0)\right)^\dagger
\mathbf{U}^n(s)}{\Delta_{nm}(0)}\mathbf{|n}(s)\rangle
\nonumber \\
&&+\mathrm{i}\hbar \mathop{\sum_{n,m=0}}_{m \neq n}
\mathrm{e}^{-\frac{\mathrm{i}}{v}\omega_m(s)}b_m(0)
\frac{\mathbf{U}^m(s)\mathbf{M}^{mn}(s)}{\Delta_{nm}(s)}
\mathbf{|n}(s)\rangle.
\end{eqnarray}
Note that $\mathbf{|\Psi}^{(1)}(0)\rangle=0$ whether or not $v\rightarrow 0$, 
as it should be, and that if we have no degeneracy we recover the results of
\cite{Rig08}.
Beginning at the ground state $|0^0(0)\rangle$ we should impose
the additional condition (\ref{initialvector}), which leads to
\begin{eqnarray}
|\mathbf{\Psi}^{(1)}(s)\rangle&=&
\mathrm{i}\hbar
\sum_{n=1}\mathrm{e}^{-\frac{\mathrm{i}}{v}\omega_0(s)}
\mathbf{J}^{0n0}(s) \mathbf{U}^0(s)\mathbf{|0}(s)\rangle
-\mathrm{i}\hbar \sum_{n=1}
\mathrm{e}^{-\frac{\mathrm{i}}{v}\omega_n(s)}
\frac{\mathbf{U}^0(0)\mathbf{M}^{0n}(0)
\left(\mathbf{U}^n(0)\right)^\dagger
\mathbf{U}^n(s)}{\Delta_{n0}(0)}\mathbf{|n}(s)\rangle
\nonumber \\
&& +\mathrm{i}\hbar \sum_{n=1}
\mathrm{e}^{-\frac{\mathrm{i}}{v}\omega_0(s)}
\frac{\mathbf{U}^0(s)\mathbf{M}^{0n}(s)}{\Delta_{n0}(s)}
\mathbf{|n}(s)\rangle.
\end{eqnarray}
Since the desired solution with the appropriate initial condition
($|0^0(0)\rangle$) is the first element of the previous vector,
$|\Psi^{(1)}(s)\rangle=[\mathbf{|\Psi}^{(1)}(s)\rangle]_0$, we get
after reversing to the usual notation
\begin{eqnarray}
|\Psi^{(1)}(s)\rangle&=&
\mathrm{i}\hbar\!
\sum_{n=1}\!\mathrm{e}^{-\frac{\mathrm{i}}{v}\omega_0(s)}
\!\!\sum_{g_0=0}\![\mathbf{J}^{0n0}(s)
\mathbf{U}^0(s)]_{0g_0}|0^{g_0}(s)\rangle
-\mathrm{i}\hbar \!\sum_{n=1}\!\!
\mathrm{e}^{-\frac{\mathrm{i}}{v}\omega_n(s)}
\!\sum_{g_n=0}\!\frac{[\mathbf{U}^0(0)\mathbf{M}^{0n}(0)
\left(\mathbf{U}^n(0)\right)^\dagger
\mathbf{U}^n(s)]_{0g_n}}{\Delta_{n0}(0)}|n^{g_n}(s)\rangle
\nonumber \\
&& +\mathrm{i}\hbar \sum_{n=1}
\mathrm{e}^{-\frac{\mathrm{i}}{v}\omega_0(s)}
\sum_{g_n=0}\frac{[\mathbf{U}^0(s)\mathbf{M}^{0n}(s)]_{0g_n}}
{\Delta_{n0}(s)}|n^{g_n}(s)\rangle. \label{first}
\end{eqnarray}
\end{widetext}

It is important to stress, as indicated in Ref.  \cite{Rig08} 
for the non-degenerate case, 
that the first term in the rhs of Eq.~(\ref{first}) is generally missing in 
standard corrections to AA  \cite{Tho83,Che11}.  
As we will see, when applying these ideas to an exactly solvable example, 
this term is also of upmost importance to obtain the correct first order correction to DAA.

\section{Conditions for the validity of the adiabatic theorem}
\label{NSC}

Now that we have developed the right tool, namely DAPT, we are able to
establish conditions for the validity of DAT, as 
stated in Sec. \ref{DAT}.  Important to assess such conditions is the 
ansatz (\ref{ansatz2}). Since the highly oscillatory terms of order 
$\mathcal{O}(1/v)$ and below appear at any order $p$, 
by properly choosing the small parameter $v$ one can always make 
DAPT perturbative series converge. Hence, the conditions that make DAPT converge
supplemented with the condition that the sum of all higher orders $p\geq 1$ is negligible
compared to the zeroth order, are sufficient conditions
to guarantee the validity of DAT. Contrariwise, if DAPT converges and the system is 
said to be well described by DAA, then the sum of all higher order terms must 
be small when compared to the zeroth order term, showing this condition is necessary too. 
In other words, if DAPT converges we can furnish rigorous necessary and sufficient conditions for
the validity of DAT. 
However, as we will see in Sec. \ref{SC}, testing for the convergence 
of DAPT series is not an easy task in general. To overcome this
limitation we develop practical necessary and sufficient conditions in the 
following sections. 

The necessary condition given in Sec. \ref{NC} 
is a generalization to degenerate systems of 
the standard quantitative condition recently proved to be necessary
for non-degenerate Hamiltonians \cite{Ton10}. On the other hand, the sufficient
condition of Sec. \ref{SC} relies heavily on DAPT and on the conditions
under which higher order terms appearing in the perturbative series of DAPT
are negligible when compared to the zeroth order.

\subsection{Necessary condition}
\label{NC}

To arrive at a necessary condition that is also practical we 
follow Tong \cite{Ton10} and others \cite{Ber84,Wil84,Wil11} and 
assume that if DAT is valid then the system is well described by DAA
and all measurements at \textit{any time}
must indeed be consistent with this assumption. 
This has a profound implication on the approximate dynamics the system obeys 
in the following sense \cite{Ton10}. 

It is not only the fidelity, $|\langle\Psi^{(0)}(t)|\Psi(t)\rangle|$, 
between the exact solution and DAA that must be close to one for the system's
dynamics to be considered adiabatic. The expectation values of any observable 
associated to the system should also be close to the ones computed with
$|\Psi^{(0)}(t)\rangle$, in particular those related to 
its geometric phase.   

Therefore, for that to be true, we must have that 
\begin{itemize}
\item[(1)] DAA approximately satisfies SE.
\item[(2)] The transition probabilities to excited \textit{eigenspaces} are negligible.
\end{itemize}

Mathematically, these two assumptions read
\begin{itemize}
\item[(1)] 
$
\mathrm{i}\,\hbar\, |\mathbf{\dot{\Psi}}^{0}(t)\rangle \approx
\mathbf{H}(t) |\mathbf{\Psi}^{0}(t)\rangle
$ and
\item[(2)]
$\left\|\mathbf{\langle n}(t)|^T\mathbf{|\Psi}(t)\rangle^T\right\|_\text{max} \ll 1,
\hspace{.5cm} n \neq 0,
$
\end{itemize}
%
%
%
%
where $\|\cdot\|_{\sf max}$ is the ``max norm'', i.e., the condition $\ll 1$ must
be tested against the absolute value of all elements of the matrix above
(the bra-column vector and the ket-row vector are combined according to the 
usual matrix multiplication rule).

From these hypotheses we can derive some lemmas that will be employed in the
forthcoming demonstration of the necessary condition.

Since DAA approximately fulfill SE (assumption 1) 
\textit{then} we may take it as a good approximation to the exact solution
$|\mathbf{\Psi}(t)\rangle$, i.e.,
\begin{equation}
|\mathbf{\Psi}(t)\rangle \approx |\mathbf{\Psi}^{(0)}(t)\rangle
\longrightarrow \text{lemma 1}.
\label{lemma1}
\end{equation}

Now, using SE, lemma 1, and assumption 1
we get
\begin{equation}
\mathrm{i}\,\hbar\, |\mathbf{\dot{\Psi}}(t)\rangle =
\mathbf{H}(t) |\mathbf{\Psi}(t)\rangle \approx
\mathbf{H}(t) |\mathbf{\Psi}^{(0)}(t)\rangle \approx
\mathrm{i}\,\hbar\, |\mathbf{\dot{\Psi}}^{(0)}(t)\rangle,
\label{prooflemma2}
\end{equation}
leading finally to lemma 2,
\begin{equation}
|\mathbf{\dot{\Psi}}(t)\rangle \approx |\mathbf{\dot{\Psi}}^{(0)}(t)\rangle
\longrightarrow \text{lemma 2}.
\label{lemma2}
\end{equation}

It is worth noting, as Tong did in the non-degenerate case \cite{Ton10}, that (\ref{lemma2}) is not a
trivial result obtained by differentiating both sides of (\ref{lemma1}). 
Equation (\ref{lemma2}) is a consequence of the fact that the state describing 
the time evolution of the system must satisfy SE, at least approximately,
which is what (\ref{prooflemma2}) is meant to show. 

Our goal next is to prove that
the quantitative condition 
\begin{equation}
\hbar\left\|\frac{\mathbf{M}^{n0}(t)}{\Delta_{n0}(t)}\right\|_1
\ll 1, \hspace{.5cm} n \neq 0, \hspace{.5cm} t \in [0,T],
\label{strongerMat}
\end{equation}
is a necessary condition for the validity of DAT. In other words, we want
to prove that (\ref{strongerMat}) follows 
from assumptions 1 and 2 (or equivalently from lemmas 1 and 2
and assumption 2). Here $\left \| A \right \|_1 = \max \limits_{1 \leq j
\leq q} \sum _{i=1}^p | a_{ij} |$ is the maximum absolute
column sum of matrix $\mathbf{A}$ with dimensions $p\times q$.

We start the proof writing the following identity for $n\neq 0$,
%
%
%
\begin{equation}
\mathbf{\langle n}(t)|^T\mathbf{|\Psi}(t)\rangle^T
= \frac{\mathbf{\langle n}(t)|^T(\mathbf{H}(t) - E_0(t))\mathbf{|\Psi}(t)\rangle^T}{\Delta_{n0}(t)}.
\end{equation}
Using SE, Eq.~(\ref{SE}), we get
\begin{eqnarray}
\mathbf{\langle n}(t)|^T&\hspace{-.4cm}|&\hspace{-.4cm}\mathbf{\Psi}(t)\rangle^T
= \frac{\mathbf{\langle n}(t)|^T\left(\mathrm{i}\hbar|\dot{\mathbf{\Psi}}(t)\rangle^T
- E_0(t)\mathbf{|\Psi}(t)\rangle^T\right)}{\Delta_{n0}(t)} \nonumber \\
&\approx&\hspace{-.2cm} \frac{\mathbf{\langle n}(t)|^T\!\!\left(\mathrm{i}\hbar|\dot{\mathbf{\Psi}}^{(0)}(t)\rangle^T
- E_0(t)\mathbf{|\Psi}^{(0)}(t)\rangle^T\right)}{\Delta_{n0}(t)},
\label{lemma1and2}
\end{eqnarray}
where the last mathematical step comes from Eqs. (\ref{lemma1})
and ({\ref{lemma2}}). Taking the transpose of ($\ref{vector00}$)
we get
\begin{equation}
\mathbf{|\Psi}^{(0)}(t)\rangle^T =
\mathrm{e}^{-\mathrm{i}\omega_0(t)}
\mathbf{|0}(t)\rangle^T\mathbf{U}^{0}(t)^T,
\label{noderivative}
\end{equation}
which leads to
\begin{eqnarray}
\mathrm{i}\hbar\mathbf{|\dot{\Psi}}^{(0)}\!(t)\!&\rangle^T& =
\mathrm{e}^{-\mathrm{i}\omega_0(t)}
\left(E_{0}(t)\mathbf{|0}(t)\rangle^T\mathbf{U}^{0}(t)^T\right.\nonumber \\
&+&\hspace{-.2cm}\left.\mathrm{i}\hbar \mathbf{|0}(t)\rangle^T\dot{\mathbf{U}}^{0}(t)^T
\hspace{-.1cm}+ \mathrm{i}\hbar|\dot{\mathbf{0}}(t)\rangle^T\mathbf{U}^{0}(t)^T
\right)\hspace{-.1cm}.
\label{derivative}
\end{eqnarray}
Inserting Eqs. (\ref{noderivative}) and (\ref{derivative}) into
(\ref{lemma1and2}) and noting that
$\mathbf{\langle n}(t)|^T\mathbf{|0}(t)\rangle^T=\mathbf{0}$ since $n\neq 0$ we
obtain
\begin{eqnarray}
\mathbf{\langle n}(t)|^T\mathbf{|\Psi}(t)\rangle^T
&\approx& \mathrm{i}\hbar\mathrm{e}^{-\mathrm{i}\omega_0(t)}
\frac{\mathbf{\langle n}(t)|^T|\dot{\mathbf{0}}(t)\rangle^T\mathbf{U}^{0}(t)^T}
{\Delta_{n0}(t)} \nonumber \\
&=&\hspace{-.1cm} \mathrm{i}\hbar\mathrm{e}^{-\mathrm{i}\omega_0(t)}
\hspace{-.05cm}\frac{\mathbf{\langle n}(t)|^T[\mathbf{U}^{0}(t)|\dot{\mathbf{0}}(t)\rangle]^T}
{\Delta_{n0}(t)}\hspace{-.05cm}.
\end{eqnarray}
Taking the max norm of both sides and using assumption $2$
we get the necessary condition
\begin{equation}
\hbar\left\|\frac{\mathbf{\langle n}(t)|^T(\mathbf{U}^0(t)\mathbf{|\dot{0}}(t)\rangle)^T}
{\Delta_{n0}(t)}\right\|_{\sf max} \ll 1, \hspace{.3cm} n \neq 0, \hspace{.3cm} t \in [0,T].
\label{nec2}
\end{equation}

In order to get a phase-free necessary condition it is convenient to 
work with (\ref{nec2}) in the standard notation. 
First note that 
\begin{eqnarray}
[\mathbf{U}^0(t)|\dot{\mathbf{0}}(t)\rangle]^T_{j_0h_0} &=&
[\mathbf{U}^0(t)|\dot{\mathbf{0}}(t)\rangle]_{h_0j_0}
\nonumber \\
&=& \sum_{g_0=0}^{d_0-1}[\mathbf{U}^0(t)]_{h_0g_0}(t)
[|\dot{\mathbf{0}}(t)\rangle]_{g_0j_0}
\nonumber \\
&=&\sum_{g_0=0}^{d_0-1}U_{h_0g_0}^0(t)|\dot{0}^{g_0}(t)\rangle,
\label{almost1}
\end{eqnarray}
with $j_0=0$, i.e., $[|\dot{\mathbf{0}}(t)\rangle]_{g_0j_0}
=[|\dot{\mathbf{0}}(t)\rangle]_{g_0}$
is a column vector and $[\mathbf{U}^0(t)|\dot{\mathbf{0}}(t)\rangle]^T_{j_0h_0}$ 
a row vector. Also, with that in mind 
\begin{eqnarray}
[\mathbf{\langle n}(t)|^T(\mathbf{U}^0(t)|\dot{\mathbf{0}}(t)\rangle)^T]_{g_nh_0}
&=&[\mathbf{\langle n}(t)|]^T_{g_n0}
[\mathbf{U}^0(t)|\dot{\mathbf{0}}(t)\rangle]^T_{0h_0}.\nonumber \\
\label{almost2}
\end{eqnarray}
Using that $[\mathbf{\langle n}(t)|]^T_{g_n0}=\langle n^{g_n}(t)|$
and inserting (\ref{almost1}) into (\ref{almost2}) we get
\begin{eqnarray}
[\mathbf{\langle n}(t)|^T(\mathbf{U}^0(t)\mathbf{|\dot{0}}(t)\rangle)^T]_{g_nh_0}
&=& \langle n^{g_n}(t)|\sum_{g_0=0}^{d_0-1}U_{h_0g_0}^0(t)
|\dot{0}^{g_0}(t)\rangle\nonumber \\
&=& \sum_{g_0=0}^{d_0-1}U_{h_0g_0}^0(t)\langle n^{g_n}(t)|\dot{0}^{g_0}(t)\rangle \nonumber \\
&=& \sum_{g_0=0}^{d_0-1}U_{h_0g_0}^0(t)M_{g_ng_0}^{n0}(t).
\label{almost3}
\end{eqnarray}
Finally, inserting Eq. (\ref{almost3}) into (\ref{nec2}) we obtain
\begin{equation}
\hbar\left|\sum_{g_0=0}^{d_0-1}U_{h_0g_0}^0(t)\frac{M_{g_ng_0}^{n0}(t)}
{\Delta_{n0}(t)}\right| \ll 1, \hspace{.3cm} n \neq 0, \hspace{.3cm} \forall g_n, h_0.
\label{final}
\end{equation}

Working with Eq.~(\ref{final}) we can finally arrive at (\ref{strongerMat})
by fully exploring the unitarity of $\mathbf{U}^n(t)$,
i.e., if we use the fact that $\left|U_{h_0g_0}^0(t)\right|\leq 1$  
we have
\begin{eqnarray}
\hbar\left|\sum_{g_0=0}^{d_0-1}U_{h_0g_0}^0(t)\frac{M_{g_ng_0}^{n0}(t)}
{\Delta_{n0}(t)}\right| &\leq&
\hbar\sum_{g_0=0}^{d_0-1}\left|U_{h_0g_0}^0(t)\frac{M_{g_ng_0}^{n0}(t)}
{\Delta_{n0}(t)}\right| \nonumber \\
&=&
\hbar\sum_{g_0=0}^{d_0-1}\left|U_{h_0g_0}^0(t)\right|
\left|\frac{M_{g_ng_0}^{n0}(t)}
{\Delta_{n0}(t)}\right| \nonumber \\
&\leq& \hbar\sum_{g_0=0}^{d_0-1}\left|\frac{M_{g_ng_0}^{n0}(t)}
{\Delta_{n0}(t)}\right| .
\end{eqnarray}
Therefore, a stronger necessary condition is
\begin{equation}
\hbar\sum_{g_0=0}^{d_0-1}\left|\frac{M_{g_ng_0}^{n0}(t)}
{\Delta_{n0}(t)}\right| \ll 1, \hspace{.5cm} n \neq 0,
\hspace{.5cm} \forall g_n, \hspace{.5cm} t \in [0,T], 
\label{strongerNec}
\end{equation}
which is exactly Eq.~(\ref{strongerMat}) if we use (\ref{M}).
Note that if Eq. (\ref{strongerNec}) holds, the weaker necessary condition
(\ref{nec2}) also holds and both reduce to the one in \cite{Ton10} when
no degeneracy is present. In such a case $\mathbf{M}^{n0}(t)$ is a $1
\times 1$ matrix leading to $[\mathbf{M}^{n0}(t)]_{00} = \langle
n(t)| \dot{0}(t)\rangle$.

One last remark. If for $n\neq m$ we take the time derivative
of the eigenvalue equation $\mathbf{H}(t)|n^{g_n}(t)\rangle$ $=$ $E_n(t)|n^{g_n}(t)\rangle$
and left multiply the result by $\langle m^{h_m}(s)|$ we get
\begin{equation}
M^{nm}_{h_ng_m}(t) = \langle n^{h_n}(t)|\mathbf{\dot{H}}(t)
|m^{g_m}(t)\rangle/\Delta_{mn}(t).
\label{condition}
\end{equation}
This last expression when inserted into (\ref{strongerNec})
indicates that the necessary condition for the validity 
of DAT is connected to the square of the gap between eigenspaces
and with the rate at which $\mathbf{H}(t)$ changes with time.

\subsection{Sufficient condition}
\label{SC}

We can write
formal rigorous sufficient conditions for the validity of DAT
by using the ratio test for ascertaining the convergence of 
DAPT series. Once the series is guaranteed to converge, additional
conditions must be applied to make DAA (DAPT zeroth order) 
the dominant term in the expansion. 

Let us start writing the ansatz (\ref{ansatz2}) in a form better suited
to the analysis that follows, 
\begin{equation}
|\mathbf{\Psi}(s)\rangle = \sum_{n=0}\sum_{p=0}^{\infty}
\mathbf{C}_{n}^{(p)}(s)|\mathbf{n}(s)\rangle,
\label{preCn}
\end{equation}
with
\begin{equation}
\mathbf{C}_{n}^{(p)}(s) =
\mathrm{e}^{-\frac{\mathrm{i}}{v}\omega_n(s)}v^p
\mathbf{B}_{n}^{(p)}(s).
\label{Cn}
\end{equation}

Note that for each $n$ we have a series involving the matrix
$\mathbf{C}_{n}^{(p)}(s)$, $p=0,1,\ldots, \infty$, where the matrix
element $[\mathbf{C}_{n}^{(p)}(s)]_{h_ng_n}$ is the probability amplitude
to order $p$ of the state $|n^{g_n}(s)\rangle$ in the expansion (\ref{preCn}).
As usual, $h_n$ handles different initial conditions and,
without loss of generality, we stick with $h_n=0$, any $n$. 
For other initial conditions one would take $h_n = 1, 2, \ldots, d_n-1$,
for any $n$. See Sec. \ref{DAPT} for details. 
Therefore, we can apply the ratio test to all
matrix elements of $\mathbf{C}_{n}^{(p)}(s)$ and test the
convergence of DAPT:
\begin{equation}
\lim_{p\rightarrow
\infty}\left|\frac{[\mathbf{C}_{n}^{(p+1)}(s)]_{0g_n}}
{[\mathbf{C}_{n}^{(p)}(s)]_{0g_n}}\right| < 1, \hspace{.5cm}
\forall n, g_n, \hspace{.5cm} s \in [0,1].
\label{ratio1}
\end{equation}
If all matrix coefficients satisfy the above condition then DAPT is
convergent. 

Inserting Eq.~(\ref{Cn}) into (\ref{ratio1}) we get
\begin{equation}
\lim_{p\rightarrow
\infty}\left|\frac{v[\mathbf{B}_{n}^{(p+1)}(s)]_{0g_n}}
{[\mathbf{B}_{n}^{(p)}(s)]_{0g_n}}\right| < 1, \hspace{.5cm}
\forall n, g_n, \hspace{.5cm} s \in [0,1],
\end{equation}
which after using (\ref{ansatzB}) becomes
\begin{equation}
\lim_{p\rightarrow \infty}\left|\frac{v\left[\sum_{m=0}
\mathrm{e}^{\frac{\mathrm{i}}{v}\omega_{m}(s)}
\mathbf{B}_{mn}^{(p+1)}(s)\right]_{0g_n}} {\left[\sum_{m=0}
\mathrm{e}^{\frac{\mathrm{i}}{v}\omega_{m}(s)}
\mathbf{B}_{mn}^{(p)}(s)\right]_{0g_n}}\right| < 1.
\label{inter}
\end{equation}

We can simplify further the previous equation by invoking the
comparison test. Let $S_{\alpha}=\sum_{p=0}^{\infty}\alpha_p$ and
$S_{\beta}=\sum_{p=0}^{\infty}\beta_p$ represent two series. Then
the comparison test says that if $S_{\beta}$ converges and
$|\alpha_p|\leq |\beta_p|$ then $S_{\alpha}$ also converges. If we
define 
$$\alpha_p = \left[\sum_{m=0}v^p
\mathrm{e}^{\frac{\mathrm{i}}{v}\omega_{m}(s)}
\mathbf{B}_{mn}^{(p)}(s)\right]_{0g_n},$$ 
which is just the element of the series
we are testing in Eq. ($\ref{inter}$), and 
$$\beta_p =
\sum_{m=0}v^p
\left|\left[\mathbf{B}_{mn}^{(p)}(s)\right]_{0g_n}\right|,$$ 
we clearly see that $|\alpha_p|\leq |\beta_p|$ and if 
\begin{equation}
\lim_{p\rightarrow \infty}\frac{v\sum_{m=0}
\left|\left[\mathbf{B}_{mn}^{(p+1)}(s)\right]_{0g_n}\right|}
{\sum_{m=0}
\left|\left[\mathbf{B}_{mn}^{(p)}(s)\right]_{0g_n}\right|} < 1,
\hspace{.2cm} \forall n, g_n, \hspace{.2cm} s \in [0,1],
\label{strongerSuf}
\end{equation}
by the comparison test DAPT also converges.

It is interesting to note that the small parameter $v$ is in
the numerator. Then, in principle, we can always make the DAPT series
converge by choosing a  small enough $v$. Of course, in
pathological Hamiltonians or in some real world experimental
realizations, we may need a really small $v=1/T$, indicating a prohibitively large $T$. 
This is an indication that this particular Hamiltonian  
cannot be made to change adiabatically when constrained by the
total execution time of the experiment. To solve this problem, 
we would need to either increase the running time of the experiment or 
build another device whose description is given by a different 
Hamiltonian, best suited for the duration of that particular
experiment.

The convergence condition (\ref{strongerSuf}) plus
\begin{eqnarray}
\left|\sum_{p=0}^{\infty}[\mathbf{C}_n^{(p+1)}(s)]_{0g_n}\right| &\ll&
\left|[\mathbf{C}_n^{(0)}(s)]_{0g_n}\right|,
\label{strongerSuf1b}
\end{eqnarray}
$\forall n, g_n, \hspace{.2cm} s \in [0,1]$
is what we call the rigorous sufficient condition.
Equation (\ref{strongerSuf1b}) guarantees that the sum of all
higher orders coefficients are negligible when compared to the zeroth order.
This condition is rather intuitive if we remember that from DAPT (see
Eq.~(\ref{ansatz2})) 
$[\mathbf{C}_n^{(p)}(s)]_{0g_n}=v^p\sum_{m=0}\mathrm{e}^{-\frac{\mathrm{i}}{v}\omega_m(s)}
\left[\mathbf{B}_{mn}^{(p)}(s)\right]_{0g_n}$ gives the $p$-th
order contribution of state $|n^{g_n}(s)\rangle$ to the exact
solution.

We can get dynamical phase-free sufficient conditions by noting that
\begin{eqnarray}
\left|\sum_{p=0}^{\infty}[\mathbf{C}_n^{(p+1)}(s)]_{0g_n}\right| &\leq&
v^{p+1}\sum_{m=0}
\left|\left[\mathbf{B}_{mn}^{(p+1)}(s)\right]_{0g_n}\right| 
\label{eqsuf1}
\end{eqnarray}
and also, by using Eq.~(\ref{B0}),
\begin{eqnarray}
\left|[\mathbf{C}_n^{(0)}(s)]_{0g_n}\right| &=&
\left|b_n(0)[\mathbf{U}^{n}(s)]_{0g_n}\right|.
\label{eqsuf2}
\end{eqnarray}
Therefore, inserting Eqs.~(\ref{eqsuf1}) and (\ref{eqsuf2}) into
(\ref{strongerSuf1b}) we obtain the following stronger sufficient 
condition \cite{footnote1},
\begin{eqnarray}
\sum_{p=0}^{\infty}\!\sum_{m=0}\!\!v^{p+1}\!\!
\left|\!\left[\mathbf{B}_{mn}^{(p+1)}(s)\right]_{0g_n}\!\right| &\ll
\left|b_n(0)[\mathbf{U}^{n}(s)]_{0g_n}\right|,\label{newsuf}
\end{eqnarray}
where we must test it for all $n$, $g_n$, and  $s \in [0,1]$.

The convergence condition (\ref{strongerSuf}) is not generally
useful in practice  since it is extremely difficult to compute the previous 
limit when $p\rightarrow \infty$. Also, in order to apply (\ref{newsuf}) we
must know all higher order corrections. We can come up, though, 
with a practical condition of convergence by looking at the ratio for a couple of
finite $p$ and truncating (\ref{newsuf}) at some finite $p$. 
For example, we can apply Eq.~(\ref{strongerSuf})
for $p=0,1$, and $2$ with the corresponding truncation of (\ref{newsuf}). 
If the previous equations
are satisfied by these $p$'s we would have the first order
contribution small compared to the zeroth order, the second small
compared to the first, and the third order small compared to the
second one.

The simplest of all practical tests consists in setting
$p=0$ in Eq.~(\ref{strongerSuf}) and truncating the sum (\ref{newsuf})
at $p=0$. 
In this case Eqs.~(\ref{strongerSuf})
and (\ref{newsuf}) collapse to the same expression.  
In other words, Eq.~(\ref{newsuf}) for $p=0$ is what we call
the practical sufficient condition and it can be cast as 
\begin{eqnarray}
v\sum_{m=0}
\left|\left[\mathbf{B}_{mn}^{(1)}(s)\right]_{0g_n}\right|
&\ll& |b_n(0)|\left|\left[\mathbf{U}^{n}(s)\right]_{0g_n}\right|,
\nonumber \\
&&\hspace{.2cm} \forall n,
g_n, \hspace{.2cm} s \in [0,1].
\end{eqnarray}

Assuming the system starts at the ground state $|0^{0}(0)\rangle$
(Eq.~\ref{initialvector}) we get
\begin{equation}
v\sum_{m=0}
\left|\left[\mathbf{B}_{m0}^{(1)}(s)\right]_{0g_0}\right| \ll
\left|\left[\mathbf{U}^{0}(s)\right]_{0g_0}\right|,
\hspace{.3cm}
\;\forall g_0, \hspace{.3cm} s \in [0,1]
\end{equation}
and
\begin{equation}
v\sum_{m=0}
\left|\left[\mathbf{B}_{mn}^{(1)}(s)\right]_{0g_n}\right| \ll 0,
\hspace{.3cm} \forall n\neq 0,\; \forall g_n, \hspace{.3cm} s \in
[0,1].
\end{equation}
Note that this last equation, and in particular the zero at the
rhs, comes from the fact that $b_{n}(0)=0$ if $n\neq 0$.
But the lhs is always positive which means it should be
zero. This is too strong a condition since in practice we may have
a very tiny contribution from excited states.
Also, it may happen that one or more $[\mathbf{U}^{0}(s)]_{0g_0}$ are zero. 
This means that
one or more of the eigenstates belonging to the ground eigenspace has, to order
zero, a null probability of being populated. Thus, for practical
purposes, we should only demand the lhs to be much
smaller than the smallest \textit{non-null} contributions coming 
from the coefficients of the zeroth order. This
guarantees that order zero, and in turn DAA,
is the dominant term when compared with the first order correction.
Hence, putting all these
pieces together the practical sufficient condition looks like
\begin{eqnarray}
v\sum_{m=0}
\left|\left[\mathbf{B}_{mn}^{(1)}(s)\right]_{0g_n}\right| &\ll
\min\limits_{\forall g_0}\!_+
\left(\left|\left[\mathbf{U}^{0}(s)\right]_{0g_0}\right|\right),
\nonumber \\
&
\hspace{.3cm} \forall n,g_n, \hspace{.3cm} s \in
[0,1], \label{suf3}
\end{eqnarray}
where $\min\!_+$ indicates that the minimum is taken over
non-null terms only.

It worth mentioning that we can increase the accuracy of the 
practical test by repeating the previous calculations for higher
orders $p$. The more orders we include the more restrictions we will
have and the stronger the sufficient test will be.
Here, we are just presenting the simplest set of conditions which, nevertheless,
turns out to be very useful as we show in Sec. \ref{example}. 

Equation (\ref{suf3}) is the practical
sufficient condition but it still depends on the geometric phase
$\mathbf{U}^{n}(s)$ at the lhs. By a similar calculation 
to the one we did when working with the necessary condition,
  we can get rid of these unitary matrices though. This procedure 
gives us a stronger and phase-free practical sufficient test.

Using Eq.~(\ref{first}) we can show that (\ref{suf3}) is equivalent to
\begin{equation}
v\hbar\left|\left[\sum_{n=1}\mathbf{J}^{0n0}(s)\mathbf{U}^0(s)
\right]_{0g_0}\right|
\ll \min\limits_{\forall g_0}\!_+
\left(\left|\left[\mathbf{U}^{0}(s)\right]_{0g_0}\right|\right)
\label{76}
\end{equation}
and
\begin{widetext}
\begin{eqnarray}
\!\!\!\!\!\frac{v\hbar}{|\Delta_{n0}(0)|}\left(
\left|\left[ \mathbf{U}^0(s)\mathbf{M}^{0n}(s)
\right]_{0g_n}\right|
\right.
\left. +\left|\left[ \mathbf{U}^0(0)\mathbf{M}^{0n}(0)
\left(\mathbf{U}^n(0)\right)^\dagger \mathbf{U}^n(s)
\right]_{0g_n}\right|
\right)
\ll \min \limits_{\forall g_0}\!_+
\left(\left|\left[\mathbf{U}^{0}(s)\right]_{0g_0}\right|\right),
%
\hspace{.05cm} \forall n\neq 0,\; \forall g_n,
\label{77}
\end{eqnarray}
%
%
where we used $|a-e^{i\varphi}b|\leq |a|+|b|$ 
to arrive at the last inequality.
Employing the Schwarz inequality we can simplify further the previous two
equations. Let us start with the lhs of the first one,
\begin{eqnarray}
\left|\left[\sum_{n=1}\mathbf{J}^{0n0}(s)
\mathbf{U}^0(s)\right]_{0g_0}\right| &\leq&
\sum_{n=1}\left|\left[\mathbf{J}^{0n0}(s)\mathbf{U}^0(s)\right]_{0g_0}\right|
=\sum_{n=1}\left|\sum_{j_0=0}^{d_0-1}\left[\mathbf{J}^{0n0}(s)\right]_{0j_0}
\left[\mathbf{U}^0(s)\right]_{j_0g_0}\right| \nonumber \\
&\leq&
\sum_{n=1}\sum_{j_0=0}^{d_0-1}\left|\left[\mathbf{J}^{0n0}(s)\right]_{0j_0}\right|
\left|\left[\mathbf{U}^0(s)\right]_{j_0g_0}\right| 
\leq
\sum_{n=1}\sum_{j_0=0}^{d_0-1}\left|\left[\mathbf{J}^{0n0}(s)\right]_{0j_0}\right|,
\label{78}
\end{eqnarray}
where the last inequality comes from the fact that
$\|\mathbf{U}^0(s)\|_{\sf max}\leq 1$. We can further
simplify the above equation looking at $\mathbf{J}^{0n0}(s)$ as
given in Eq.~(\ref{J}).
Working with the matrix element $[\mathbf{J}^{0n0}(s)]_{0j_0}$ we
have
%
%
\begin{eqnarray}
|[\mathbf{J}^{0n0}(s)]_{0j_0}| &=&\left|\int_0^s\mathrm{d}s'
\left(\frac{[\mathbf{U}^0(s')\mathbf{M}^{0n}(s')\mathbf{M}^{n0}(s')
\left(\mathbf{U}^0(s')\right)^\dagger]_{0j_0}}{\Delta_{0n}(s')}\right)\right|
\leq \int_0^s\mathrm{d}s'
\left(\frac{\left|[\mathbf{U}^0(s')\mathbf{M}^{0n}(s')\mathbf{M}^{n0}(s')
\left(\mathbf{U}^0(s')\right)^\dagger]_{0j_0}\right|}{|\Delta_{0n}(s')|}\right)
\nonumber \\
&=& \int_0^s\mathrm{d}s'
\left(\frac{\sum_{k_0,l_n,i_0=0}^{d_0-1}|[\mathbf{U}^0(s')]_{0k_0}[\mathbf{M}^{0n}(s')]_{k_0l_n}
[\mathbf{M}^{n0}(s')]_{l_ni_0}
[\left(\mathbf{U}^0(s')\right)^\dagger]_{i_0j_0}|}{|\Delta_{0n}(s')|}\right)
\nonumber \\
&\leq& \int_0^s\mathrm{d}s'
\left(\frac{\sum_{k_0,l_n,i_0=0}^{d_0-1}|[\mathbf{M}^{0n}(s')]_{k_0l_n}
[\mathbf{M}^{n0}(s')]_{l_ni_0}| }{|\Delta_{0n}(s')|}\right)
= \int_0^s\!\!\mathrm{d}s'\!	
\left(\frac{\sum_{k_0,i_0=0}^{d_0-1}|[\mathbf{M}^{0n}(s')
(\mathbf{M}^{0n}(s'))^\dagger]_{k_0i_0}| }{|\Delta_{0n}(s')|}\right)\!,
\label{74}
\end{eqnarray}
in which the last equation comes from the fact that 
$[\mathbf{M}^{nm}(s)]^\dagger=-[\mathbf{M}^{mn}(s)].$
Returning to (\ref{78})  with the aid of (\ref{74})
we get after noting that its rhs does not depend on $j_0$,
\begin{eqnarray}
\left|\left[\sum_{n=1}\mathbf{J}^{0n0}(s)\mathbf{U}^0(s)
\right]_{0g_0}\right|
\leq
\sum_{n=1}\sum_{j_0=0}^{d_0-1}\left|\left[\mathbf{J}^{0n0}(s)\right]_{0j_0}\right|
= d_0 \int_0^s\mathrm{d}s'\sum_{n=1}
\left(\frac{\sum_{k_0,i_0=0}^{d_0-1}|[\mathbf{M}^{0n}(s')
(\mathbf{M}^{0n}(s'))^\dagger]_{k_0i_0}|}{|\Delta_{0n}(s')|}\right).
\end{eqnarray}
Hence the first piece of our practical sufficient condition, Eq. (\ref{76}), becomes
\begin{equation}
v\hbar d_0 \int_0^s\mathrm{d}s'\sum_{n=1}
\left(\frac{\sum_{k_0,i_0=0}^{d_0-1}|[\mathbf{M}^{0n}(s')
(\mathbf{M}^{0n}(s'))^\dagger]_{k_0i_0}|}{|\Delta_{0n}(s')|}\right)
\ll \min\limits_{\forall g_0}\!_+
\left(\left|\left[\mathbf{U}^{0}(s)\right]_{0g_0}\right|\right), 
\hspace{.3cm} s \in [0,1].
\label{altSuf1a}
\end{equation}

For the the second piece of the
practical sufficient condition, Eq. (\ref{77}),
it straightforwardly follows that
\begin{eqnarray}
\left|\left[ \mathbf{U}^0(s)\mathbf{M}^{0n}(s)
\right]_{0g_n}\right| \leq
\sum_{k_0=0}^{d_0-1} \left|[\mathbf{U}^0(s)]_{0k_0}[\mathbf{M}^{0n}(s)]_{k_0g_n}
\right| 
\leq
\sum_{k_0=0}^{d_0-1} \left|[\mathbf{M}^{0n}(s)]_{k_0g_n}\right|
\label{77a}
\end{eqnarray}
and
\begin{eqnarray}
\left|\left[ \mathbf{U}^0(0)\mathbf{M}^{0n}(0)
\left(\mathbf{U}^n(0)\right)^\dagger \mathbf{U}^n(s)
\right]_{0g_n}\right| &\leq& \sum_{k_0,l_n,i_n=0}^{d_0-1,d_n-1}
\left|[\mathbf{U}^0(0)]_{0k_0}
[\mathbf{M}^{0n}(0)]_{k_0l_n}[\left(\mathbf{U}^n(0)\right)^\dagger]_{l_ni_n}
[\mathbf{U}^n(s)]_{i_ng_n} \right|\nonumber \\
&\leq& d_n\sum_{k_0,l_n=0}^{d_0-1,d_n-1} \left|
[\mathbf{M}^{0n}(0)]_{k_0l_n} \right|.
\label{78a}
\end{eqnarray}
Using Eqs.~(\ref{77a}) and (\ref{78a}), Eq.~(\ref{77}) becomes
\begin{eqnarray}
\!\!\!\!\frac{v\hbar}{|\Delta_{n0}(0)|}\left( \sum_{k_0=0}^{d_0-1}
\left|[\mathbf{M}^{0n}(s)]_{k_0g_n}\right| +
d_n\sum_{k_0,l_n=0}^{d_0-1,d_n-1} \left|
[\mathbf{M}^{0n}(0)]_{k_0l_n} \right|\right)&\hspace{-.1cm}\ll\hspace{-.1cm}& \min
\limits_{\forall g_0}\!_+
\left(\left|\left[\mathbf{U}^{0}(s)\right]_{0g_0}\right|\right),
\hspace{.1cm} \forall n\neq 0,\; \forall g_n,
\; s \in [0,1]. 
\label{altSuf1b}
\end{eqnarray}

It is worth noting that the degeneracy level of the eigenspaces are
relevant since in Eq. (\ref{altSuf1a}) we have $d_0$ and in (\ref{altSuf1b})
$d_n$ explicitly appearing in those equations. They also depend implicitly on
the degeneracy level of the eigenspaces since the sums that remain
to be computed will have more or less terms whether we have a
higher or lower degree of degeneracy. We should also remark that the
``history'' of the time evolution of the state (integration over time)
is important in the sufficient condition (\ref{altSuf1a}) while the history
does not show up in the necessary condition.

Finally, the practical sufficient conditions (\ref{altSuf1a}) and (\ref{altSuf1b})
can be expressed in terms of the real time variable $t$ as,
\begin{equation}
\hbar d_0 \int_0^t\mathrm{d}t'\sum_{n=1}
\left(\frac{\sum_{k_0,i_0=0}^{d_0-1}|[\mathbf{M}^{0n}(t')
(\mathbf{M}^{0n}(t'))^\dagger]_{k_0i_0}|}{|\Delta_{0n}(t')|}\right)
\ll \min \limits_{\forall g_0}\!_+
\left(\left|\left[\mathbf{U}^{0}(t)\right]_{0g_0}\right|\right),
\hspace{.5cm} 
\hspace{.5cm} t \in [0,T]. \label{suf1at}
\end{equation}
and
\begin{eqnarray}
\frac{\hbar}{|\Delta_{n0}(0)|}\left(\sum_{k_0=0}^{d_0-1}
\hspace{-.15cm}\left|[\mathbf{M}^{0n}(t)]_{k_0g_n}\!\right| +
d_n\hspace{-.5cm}\sum_{k_0,l_n=0}^{d_0-1,d_n-1}\hspace{-.5cm}
\left| [\mathbf{M}^{0n}(0)]_{k_0l_n} \right|\right)
&\ll& \min \limits_{\forall g_0}\!_+
\left(\left|\left[\mathbf{U}^{0}(t)\right]_{0g_0}\right|\right),\;
\forall n\neq 0,\; \forall g_n, \; t \in [0,T].
\label{suf1bt}
\end{eqnarray}

\end{widetext}

\section{An analytical  example}
\label{example}

An important aspect of any useful perturbation theory is that it should 
work when applied to simple problems whose exact solutions are known. 
It is decisive that the zeroth and higher order perturbative corrections 
exactly match 
the expansion of the exact solution in terms of the perturbative parameter.
Indeed, any error in this matching is an indication that a given
perturbation theory is prone to failure when applied to more sophisticated
problems.  It is our next goal to apply DAPT to an exactly solvable degenerate problem
and compare its zeroth and first order terms with the equivalent ones
obtained by expanding the exact solution about the perturbative parameter.
As we will see, we obtain perfect correspondence between the expansion of
the exact solution and DAPT perturbative terms. 

In order to test DAPT, and the necessary and sufficient conditions of 
Sec. \ref{NSC}, we start by presenting for the first time
all the details of the calculations that led
us to exactly solve \cite{Rig10} the time-dependent SE for the 
degenerate Hamiltonian of \cite{Bis89}.
This model is the simplest degenerate version of the exactly solvable
time-dependent non-degenerate spin-1/2 system subjected to a classical
rotating magnetic field about a fixed axis \cite{Rab54}.

\subsection{The exact solution}

Let us consider a four-level system subjected to a rotating
classical magnetic field $\mathbf{B}(t) = B \mathbf{r}(t)$ whose
magnitude is constant and given by $B =|\mathbf{B}|$. In spherical
coordinates $ \mathbf{r}(t) = (\sin \theta \cos \varphi(t), \sin
\theta \sin \varphi(t), \cos \theta ), $ with $0 \leq \theta \leq
\pi$ and $0 \leq \varphi <2\pi$ being the polar and azimuthal
angles, respectively. The Hamiltonian describing the four-level
system is \cite{Bis89}
\begin{equation}
\mathbf{H}(t) = \frac{\hbar}{2} b\, \mathbf{r}(t) \cdot \mathbf{\Gamma},
\label{Hb}
\end{equation}
where $b>0$ is proportional to the field $B$
and $\bm{\Gamma}=(\Gamma_x,\Gamma_y,\Gamma_z)$ are the Dirac matrices
$\Gamma_j = \sigma_x \otimes \sigma_j$, $j=x,y,z$. Here $\sigma_j$ are the
usual Pauli matrices, inducing the following algebra for $\Gamma_j$,
$$
\{\Gamma_i,\Gamma_j\} = 2\delta_{ij}\bm{I}_4,
$$
$$
[\Gamma_i,\Gamma_j] = 2\mathrm{i}\epsilon_{ijk}\Pi_k,
$$
where $\bm{I}_4$ is the identity matrix of dimension four,
$\delta_{ij}$ the Kronecker delta,
$\epsilon_{ijk}$ the Levi-Civita symbol, and
$\Pi_k = \bm{I}_2\otimes \sigma_k$.

 Hamiltonian (\ref{Hb}) may represent a single four-level particle coupled
to a rotating magnetic field with coupling constant $b\hbar/2$ or two
interacting spin-1/2 particles since $\Gamma_j=\sigma_x \otimes \sigma_j$.
In the latter case we have three types of interactions between the two
particles, for $j=1,2,3$, whose coupling constants are proportional to the
components of $b\mathbf{r}(t)\hbar/2$.
In the basis where $\Pi_z$ is diagonal,
$\{|\!\!\uparrow\uparrow\rangle,|\!\!\uparrow\downarrow\rangle,
|\!\!\downarrow\uparrow\rangle,|\!\!\downarrow\downarrow\rangle\}$,
the snapshot eigenvectors of $\mathbf{H}(t)$, Eq. (\ref{Hb}), are
\begin{eqnarray}
|0^0(t)\rangle\! &=&\! \frac{1}{\sqrt{2}}\!\left(\mathrm{e}^{-\mathrm{i}\varphi(t)}\sin\theta
|\!\!\uparrow \uparrow \rangle
\!-\!\cos\theta|\!\!\uparrow\downarrow\rangle \!-\!|\!\!\downarrow\downarrow\rangle \right)\!\!,
\label{zerozerostate}\\
|0^1(t)\rangle\! &=&\! \frac{1}{\sqrt{2}}\!\left(\cos\theta
| \!\!\uparrow \uparrow \rangle
+\mathrm{e}^{\mathrm{i}\varphi(t)} \sin\theta|\!\!\uparrow\downarrow\rangle
-|\!\!\downarrow \uparrow \rangle \right)\!\!,
\label{zeroonestate} \\
|1^0(t)\rangle\! &=&\! \frac{1}{\sqrt{2}}\!\left(\mathrm{e}^{-\mathrm{i}\varphi(t)}\sin\theta
|\!\!\uparrow \uparrow \rangle
\!-\!\cos\theta|\!\!\uparrow\downarrow\rangle
\!+\!|\!\!\downarrow\downarrow\rangle\right)\!\!,
\label{onezerostate}\\
|1^1(t)\rangle\! &=&\! \frac{1}{\sqrt{2}}\!\left(\cos\theta | \!\!\uparrow
\uparrow \rangle +\mathrm{e}^{\mathrm{i}\varphi(t)}
\sin\theta|\!\!\uparrow\downarrow\rangle +|\!\!\downarrow \uparrow
\rangle \right)\!\!. \label{oneonestate}
\end{eqnarray}
If we deal with a two spin-1/2 system we also have, for instance,
$|\!\!\uparrow\downarrow\rangle = |\!\uparrow\rangle\otimes|\!\downarrow\rangle$, where
$|\!\uparrow\rangle$ and $|\!\downarrow\rangle$ are the eigenstates of $\sigma_z$.
The first two eigenvectors are degenerate with energy $E_0$ while the last
two have $E_1$,
\begin{eqnarray}
E_0=-(\hbar/2) b \hspace{.3cm}&\mbox{and}&\hspace{.3cm} E_1=(\hbar/2) b \label{energy01},
\end{eqnarray}
resulting in a constant gap between the two eigenspaces, $\Delta_{10}(s)$
$=$ $\hbar b$.

If $\varphi(t) = w\,t$, with $w>0$ being the frequency of the rotating
magnetic field, we can solve the time-dependent problem exactly by employing
techniques similar to those developed for the single spin-1/2 problem
\cite{Boh93,Rab54}. We first define the rotated
state
\begin{equation}
|\tilde{\Psi}(t)\rangle = \mathcal{U}^{\dagger}(t)|\Psi(t)\rangle,
\label{psibar}
\end{equation}
with
\begin{equation}
\mathcal{U}(t) = \mathrm{e}^{-\mathrm{i}\frac{wt}{2}\Pi_z}.
\label{Urotate}
\end{equation}
Inserting Eq.~(\ref{psibar}) into the SE
one can show that $|\tilde{\Psi}(t)\rangle$ evolves according to (\ref{SE}) with a new
Hamiltonian 
\begin{equation}
\tilde{\mathbf{H}} = \mathcal{U}^{\dagger}(t)\mathbf{H}(t)\mathcal{U}(t)
- \mathrm{i}\hbar \mathcal{U}^{\dagger}(t)\frac{{\rm
d}\mathcal{U}(t)}{{\rm d} t}.
\label{Hbar}
\end{equation}
Using Eq.~(\ref{Urotate}) and the mathematical identity
$$
\Gamma_x \cos(wt) + \Gamma_y \sin(wt) =
\mathcal{U}(t)\Gamma_x\mathcal{U}^{\dagger}(t),
$$
that results from the fact that
$[\Pi_z,\Gamma_x] = \mathrm{i}2\Gamma_y$ and
$[\Pi_z,\Gamma_y] = -\mathrm{i}2\Gamma_x$,
one can show that Eq.~(\ref{Hbar}) can be written as
\begin{eqnarray}
\tilde{\mathbf{H}} &=& \frac{\hbar}{2}\left(
b\Gamma_x\sin\theta +b\Gamma_z\cos\theta - w\Pi_z
\right).
\label{Hbar2}
\end{eqnarray}
Since $\tilde{\mathbf{H}}$ is time
independent
$
|\tilde{\Psi}(t)\rangle=\mathrm{e}^{-\mathrm{i}\frac{
\tilde{\mathbf{H}}t}{\hbar}}
|\tilde{\Psi}(0)\rangle.
$
Inverting Eq.~(\ref{psibar}) and noting that
$|\tilde{\Psi}(0)\rangle =|\Psi(0)\rangle$ the solution to the
original problem is
\begin{equation}
|\Psi(t)\rangle = \mathrm{e}^{-\mathrm{i}\frac{wt}{2}\Pi_z}
\mathrm{e}^{-\mathrm{i}\frac{\tilde{\mathbf{H}}t}{\hbar}}
|\Psi(0)\rangle.
\label{psit}
\end{equation}

We need now to rewrite the general solution (\ref{psit})
in terms of the snapshot eigenvectors (\ref{zerozerostate})-(\ref{oneonestate}).
For ease of notation, we
define the following vectors:
\begin{eqnarray}
\mathbf{w} &=& w \mathbf{z}, \\
\mathbf{b} &=& b \mathbf{r}(t), \\
\mathbf{\Omega_\pm} &=& \mathbf{w} \pm \mathbf{b},
\end{eqnarray}
where $\mathbf{z}$ is the unity vector parallel to the
$z$-direction and $\theta$ the angle between $\mathbf{w}$ and
$\mathbf{b}$. This gives for the magnitude of
$\mathbf{\Omega_\pm}$,
\begin{equation}
\Omega^2_\pm = w^2 + b^2 \pm 2wb\cos\theta. \label{Omega}
\end{equation}
The eigenvalues of $\tilde{\mathbf{H}}$ are
\begin{equation}
\tilde{E}_1 = -\frac{\hbar}{2}\Omega_-, \hspace{.2cm} \tilde{E}_2
= \frac{\hbar}{2}\Omega_-, \hspace{.2cm} \tilde{E}_3 = -
\frac{\hbar}{2}\Omega_+, \hspace{.2cm} \tilde{E}_4 =
\frac{\hbar}{2}\Omega_+.
\end{equation}
Note that the transformed Hamiltonian is no longer degenerate. Its
eigenvectors are respectively
\begin{eqnarray*}
|\tilde{E}_1\rangle\! &=&\!
\frac{\tilde{\Omega}_{+-}}{2\sqrt{\Omega_-}}\!\left (\!
|\!\!\uparrow\uparrow\rangle
-\frac{b\sin\theta}{\tilde{\Omega}_{+-}^2}|\!\!\uparrow\downarrow\rangle
+ |\!\!\downarrow\uparrow\rangle
-\frac{b\sin\theta}{\tilde{\Omega}_{+-}^2}
|\!\!\downarrow\downarrow\rangle\!\right)\!\!, \\
|\tilde{E}_2\rangle\! &=&\!
\frac{\tilde{\Omega}_{-+}}{2\sqrt{\Omega_-}}\!\left(\!
|\!\!\uparrow\uparrow\rangle
+\frac{b\sin\theta}{\tilde{\Omega}_{-+}^2}|\!\!\uparrow\downarrow\rangle
+ |\!\!\downarrow\uparrow\rangle
+\frac{b\sin\theta}{\tilde{\Omega}_{-+}^2}
|\!\!\downarrow\downarrow\rangle\!\right)\!\!,\\
|\tilde{E}_3\rangle\! &=&\!
\frac{\tilde{\Omega}_{++}}{2\sqrt{\Omega_+}}\!\left(\!
|\!\!\uparrow\uparrow\rangle
+\frac{b\sin\theta}{\tilde{\Omega}_{++}^2}|\!\!\uparrow\downarrow\rangle
- |\!\!\downarrow\uparrow\rangle
-\frac{b\sin\theta}{\tilde{\Omega}_{++}^2}
|\!\!\downarrow\downarrow\rangle\!\right)\!\!,\\
|\tilde{E}_4\rangle\! &=&\!
\frac{\tilde{\Omega}_{--}}{2\sqrt{\Omega_+}}\!\left(\!
|\!\!\uparrow\uparrow\rangle
-\frac{b\sin\theta}{\tilde{\Omega}_{--}^2}|\!\!\uparrow\downarrow\rangle
- |\!\!\downarrow\uparrow\rangle
+\frac{b\sin\theta}{\tilde{\Omega}_{--}^2}
|\!\!\downarrow\downarrow\rangle\!\right)\!\!,
\end{eqnarray*}
where we have defined
\begin{eqnarray*}
\tilde{\Omega}_{--}\! =\! \sqrt{\Omega_+ - w - b\cos\theta}, &
\tilde{\Omega}_{++} \!=\! \sqrt{\Omega_+ + w +
b\cos\theta}, \nonumber \\
\tilde{\Omega}_{-+} \!=\! \sqrt{\Omega_- - w + b\cos\theta}, &
\tilde{\Omega}_{+-} \!=\! \sqrt{\Omega_- + w - b\cos\theta}.
\end{eqnarray*}

A general initial state $|\Psi(0)\rangle$ can be written as
\begin{equation}
|\Psi(0)\rangle = \sum_{j=1}^{4}\tilde{a}_j(0)|\tilde{E}_j\rangle,
\label{tilde0}
\end{equation}
with $\tilde{a}_j(0)$ given by the initial conditions. Inserting
Eq.~(\ref{tilde0})  into (\ref{psit}) we get,
\begin{equation}
|\Psi(t)\rangle = \mathrm{e}^{-\mathrm{i}\frac{wt}{2}\Pi_z}
\sum_{j=1}^{4}\mathrm{e}^{-\mathrm{i}\frac{\tilde{E}_jt}{\hbar}}\tilde{a}_j(0)
|\tilde{E}_j\rangle. \label{psitt}
\end{equation}

By looking at the
definition of $\Pi_z$ it is not difficult to see that
$\Pi_z|\!\!\uparrow\uparrow\rangle =|\!\!\uparrow\uparrow\rangle$,
$\Pi_z|\!\!\uparrow\downarrow\rangle =-|\!\!\uparrow\downarrow\rangle$,
$\Pi_z|\!\!\downarrow\uparrow\rangle =|\!\!\downarrow\uparrow\rangle$, and
$\Pi_z|\!\!\downarrow\downarrow\rangle
=-|\!\!\downarrow\downarrow\rangle$. Thus, $|\Psi(t)\rangle$ becomes
\begin{equation}
|\Psi(t)\rangle =
\sum_{j=1}^{4}\mathrm{e}^{-\mathrm{i}\frac{\tilde{E}_jt}{\hbar}}\tilde{a}_j(0)
|\tilde{F}_j\rangle, \label{psittt}
\end{equation}
where
\begin{eqnarray*}
|\tilde{F}_1\rangle &=& \frac{\mathrm{e}^{-\mathrm{i}\frac{wt}{2}}
\tilde{\Omega}_{+-}}{2\sqrt{\Omega_-}}\left(
|\!\!\uparrow\uparrow\rangle + |\!\!\downarrow\uparrow\rangle
-\frac{\mathrm{e}^{\mathrm{i}wt}b\sin\theta}{\tilde{\Omega}_{+-}^2}
(|\!\!\uparrow\downarrow\rangle + |\!\!\downarrow\downarrow\rangle)\right), \\
|\tilde{F}_2\rangle &=& \frac{\mathrm{e}^{-\mathrm{i}\frac{wt}{2}}
\tilde{\Omega}_{-+}}{2\sqrt{\Omega_-}}\left(
|\!\!\uparrow\uparrow\rangle + |\!\!\downarrow\uparrow\rangle
+\frac{\mathrm{e}^{\mathrm{i}wt}b\sin\theta}{\tilde{\Omega}_{-+}^2}
(|\!\!\uparrow\downarrow\rangle + |\!\!\downarrow\downarrow\rangle)\right),\\
|\tilde{F}_3\rangle &=& \frac{\mathrm{e}^{-\mathrm{i}\frac{wt}{2}}
\tilde{\Omega}_{++}}{2\sqrt{\Omega_+}}\left(
|\!\!\uparrow\uparrow\rangle - |\!\!\downarrow\uparrow\rangle +
\frac{\mathrm{e}^{\mathrm{i}wt}b\sin\theta}{\tilde{\Omega}_{++}^2}
(|\!\!\uparrow\downarrow\rangle - |\!\!\downarrow\downarrow\rangle)\right),\\
|\tilde{F}_4\rangle &=& \frac{\mathrm{e}^{-\mathrm{i}\frac{wt}{2}}
\tilde{\Omega}_{--}}{2\sqrt{\Omega_+}}\left(
|\!\!\uparrow\uparrow\rangle - |\!\!\downarrow\uparrow\rangle
-\frac{\mathrm{e}^{\mathrm{i}wt}b\sin\theta}{\tilde{\Omega}_{--}^2}
(|\!\!\uparrow\downarrow\rangle -
|\!\!\downarrow\downarrow\rangle)\right).
\end{eqnarray*}

Since we want to compare the exact solution with DAPT 
when the system starts at the ground state $|0^{0}(0)\rangle$,
we need to determine 
$\tilde{a}_j(0)$ for
\begin{equation}
|\Psi(0)\rangle=|0^0(0)\rangle = \frac{1}{\sqrt{2}}\left(\sin\theta |\!\!\uparrow
\uparrow \rangle -\cos\theta|\!\!\uparrow\downarrow\rangle
-|\!\!\downarrow\downarrow\rangle\right). \label{initialpsi}
\end{equation}
Equating Eq.~(\ref{initialpsi}) with (\ref{psittt}) at $t=0$ we
get a linear system of four equations in the variables
$\tilde{a}_j(0)$, $j=1,\ldots, 4$, whose solution is,
\begin{eqnarray}
\tilde{a}_1(0) &=& \frac{(\Omega_- - w + b) \tilde{\Omega}_{+-}
}{2b\sqrt{2\Omega_-}}\cot(\theta/2), \label{a1}\\
\tilde{a}_2(0) &=& -\frac{(\Omega_- + w - b) \tilde{\Omega}_{-+}
}{2b\sqrt{2\Omega_-}}\cot(\theta/2),\\
\tilde{a}_3(0) &=& \frac{(\Omega_+ - w + b) \tilde{\Omega}_{++}
}{2b\sqrt{2\Omega_+}}\tan(\theta/2),\\
\tilde{a}_4(0) &=&- \frac{(\Omega_+ + w - b) \tilde{\Omega}_{--}
}{2b\sqrt{2\Omega_+}}\tan(\theta/2) \label{a4}.
\end{eqnarray}

Now that we have the exact solution, with the appropriate initial
condition, we just need to re-write it in terms of the snapshot
eigenvectors of the original Hamiltonian. In this way, one 
can straightforwardly compare  the
expansion of the exact solution up to first order with the
perturbative corrections coming from DAPT. Inverting
Eqs.~(\ref{zerozerostate})-(\ref{oneonestate}) we get
\begin{eqnarray}
|\!\!\uparrow\uparrow\rangle &=& \frac{1}{\sqrt{2}} \left(
\mathrm{e}^{\mathrm{i}wt}\sin\theta|0^0(t)\rangle + \cos\theta |0^1(t)\rangle 
\right.\nonumber \\
&&+\left.
\mathrm{e}^{\mathrm{i}wt}\sin\theta|1^0(t)\rangle + \cos\theta|1^1(t)\rangle
\right), \label{upup}\\
|\!\!\uparrow\downarrow\rangle &=& \frac{1}{\sqrt{2}} \left(
-\cos\theta|0^0(t)\rangle +\mathrm{e}^{-\mathrm{i}wt}\sin\theta |0^1(t)\rangle
\right. \nonumber \\ 
&&\left.-
\cos\theta|1^0(t)\rangle +\mathrm{e}^{-\mathrm{i}wt}\sin\theta|1^1(t)\rangle
\right),\\
|\!\!\downarrow\uparrow\rangle &=& \frac{1}{\sqrt{2}} \left( -
|0^1(t)\rangle + |1^1(t)\rangle
\right),\\
|\!\!\downarrow\downarrow\rangle &=& \frac{1}{\sqrt{2}} \left( -
|0^0(t)\rangle + |1^0(t)\rangle \right). \label{downdown}
\end{eqnarray}
Inserting Eqs.~(\ref{upup})-(\ref{downdown}) into (\ref{psittt})
and using the initial condition (\ref{a1})-(\ref{a4}) we get after
a long but straightforward algebraic manipulation
\begin{eqnarray}
|\Psi(t)\rangle \!&=&\! \frac{\mathrm{e}^{\mathrm{i} w t/2}}{2}\!\left[\! (1\! +\! \cos\theta)
A_-(t) \!+\! (1\!-\!\cos\theta)A_+(t)\!\right]\! |0^0(t)\rangle \nonumber \\
&& + \frac{\mathrm{e}^{-\mathrm{i}w t/2}\sin\theta}{2}\left( A_+(t) -
A_-(t)\right) |0^1(t)\rangle \nonumber \\
&& + \frac{\mathrm{e}^{\mathrm{i}w t/2}\sin^2\theta}{2} \left( B_+(t) +
B_-(t)\right) |1^0(t)\rangle \nonumber \\
&& + \frac{\mathrm{e}^{-\mathrm{i} w t/2}}{2} \sin\!\theta\! \left[\! (1 \!+\! \cos\theta)
B_-(t)\right.\nonumber \\
&&\left.- (1\!-\! \cos\!\theta) B_+(t)\! \right]\!|1^1(t)\rangle,
\label{ground_exact}
\end{eqnarray}
where
\begin{eqnarray}
A_\pm(t) \hspace{-.15cm}&=&\hspace{-.15cm} \cos\!\left(\Omega_\pm t/2\right) + \mathrm{i}\frac{b\pm
w\cos\!\theta}{\Omega_\pm}\sin\!\left(\Omega_\pm t/2\right)\!, \label{Apm} \\
B_\pm(t) &=& \mathrm{i}\frac{w}{\Omega_\pm}\sin(\Omega_\pm t /2).
\label{Bpm}
\end{eqnarray}

\subsection{Expansion of the exact solution}

We can write Eq.~(\ref{ground_exact}) as
\begin{eqnarray}
|\Psi(t)\rangle &=& c_{00}(t) |0^0(t)\rangle + c_{01}(t)
|0^1(t)\rangle + c_{10}(t) |1^0(t)\rangle \nonumber \\
&&+ c_{11}(t)
|1^1(t)\rangle,
\label{cij}
\end{eqnarray}
and our goal is to expand each one of $c_{ij}(t)$ up to first order. 
As in the non-degenerate case \cite{Rig08} 
we choose $v=w$ for the small perturbative parameter, a natural
choice since we have experimental control over 
the frequency of the rotating field. The smaller $w$ the better is DAA
approximation to the exact solution. Also, we must be careful while expanding the exact 
solution \cite {Rig08}
since terms $w^{n+1} t$ are of order $w^n$ because $t=s/v=s/w$.

\subsubsection{Expansion of $c_{00}(t)$}

Looking at Eq.~(\ref{ground_exact}) we see that we first need to
expand $A_\pm(t)$, Eq.~(\ref{Apm}), up to first order. Using
\begin{displaymath}
\frac{b \pm w\cos\theta}{\Omega_\pm} = 1 + \mathcal{O}(w^2)
\end{displaymath}
we have
\begin{equation}
A_\pm(t) = \mathrm{e}^{\mathrm{i}\frac{\Omega_\pm t}{2}} +
\mathcal{O}(w^2),
\label{Apm2}
\end{equation}
which leads to
\begin{eqnarray*}
c_{00}(t) &=& \frac{1}{2}\left[\mathrm{e}^{\mathrm{i}(w+\Omega_-) t/2}(1 +
\cos\theta) \right.\nonumber \\
 &&\left. +\mathrm{e}^{\mathrm{i} (w+\Omega_+) t/2}(1-\cos\theta))\right] +
\mathcal{O}(w^2).
\end{eqnarray*}
Let us now look at the first exponential. Since
\begin{displaymath}
w+\Omega_- = b + (1-\cos\theta)w + \frac{w^2}{2b}\sin^2\theta +
\mathcal{O}(w^3)
\end{displaymath}
we obtain
\begin{eqnarray}
\mathrm{e}^{\mathrm{i} (w+\Omega_-) t/2} &=&\mathrm{e}^{\mathrm{i} b t/2}\mathrm{e}^{\mathrm{i}(1-\cos\theta)wt/2}\left(
1 + \mathrm{i}\frac{w^2t}{4b}\sin^2\theta\right)\nonumber \\
 && +\mathcal{O}(w^2). \label{expOminus}
\end{eqnarray}
A similar analysis leads to
\begin{eqnarray}
\mathrm{e}^{\mathrm{i} (w+\Omega_+) t/2} &=&\mathrm{e}^{\mathrm{i} b t/2}\mathrm{e}^{\mathrm{i}(1+\cos\theta)wt/2}\left(
1 + \mathrm{i}\frac{w^2t}{4b}\sin^2\theta\right)\nonumber \\
&& + \mathcal{O}(w^2). \label{expOplus}
\end{eqnarray}
Combining both results we get after a little algebra
\begin{eqnarray}
c_{00}(t) &=&\mathrm{e}^{\mathrm{i} \frac{b t}{2}}\mathrm{e}^{\mathrm{i} \frac{w t}{2}} \left( 1 +
\mathrm{i}\frac{w^2t}{4b}\sin^2\theta\right)\left[\cos\left(\frac{wt}{2}\cos\theta\right)
\right. \nonumber \\
&&\left.- \mathrm{i}\cos\theta\sin\left(\frac{wt}{2}\cos\theta\right)\right] +
\mathcal{O}(w^2). 
\label{c00}
\end{eqnarray}

\subsubsection{Expansion of $c_{01}(t)$}

Similarly,
\begin{equation}
c_{01}(t) =\frac{1}{2}\mathrm{e}^{-\mathrm{i} \frac{w t}{2}} \sin\theta \left(\mathrm{e}^{\mathrm{i}
\Omega_+ t/2} -\mathrm{e}^{\mathrm{i} \Omega_- t/2}\right) + \mathcal{O}(w^2).
\end{equation}
Using Eqs.~(\ref{expOminus}) and (\ref{expOplus}) we get
after some algebra
\begin{eqnarray}
c_{01}(t) &=& \mathrm{i}\mathrm{e}^{\mathrm{i} \frac{b t}{2}}\mathrm{e}^{-\mathrm{i} \frac{w t}{2}} \left( 1 +
\mathrm{i}\frac{w^2t}{4b}\sin^2\theta\right)\sin\theta\sin\left(\frac{wt}{2}\cos\theta\right)
\nonumber \\
&&+ \mathcal{O}(w^2). \label{c01}
\end{eqnarray}

\subsubsection{Expansion of $c_{10}(t)$}

In this case we need
\begin{eqnarray}
B_\pm(t) &=& \mathrm{i} \frac{w}{b}\sin\left(\frac{\Omega_\pm t}{2}\right)+
\mathcal{O}(w^2) \nonumber \\ 
&=& \frac{w}{2b}\mathrm{e}^{\mathrm{i}\frac{\Omega_\pm
t}{2}}(1-\mathrm{e}^{-\mathrm{i}\Omega_\pm t})+ \mathcal{O}(w^2).
\end{eqnarray}
Expanding the term inside the parenthesis up to zeroth order
gives
\begin{equation}
B_\pm(t) = \frac{w}{2b}\mathrm{e}^{\mathrm{i}\frac{\Omega_\pm t}{2}}(1-\mathrm{e}^{-\mathrm{i}b
t}\mathrm{e}^{\mp \mathrm{i}w t \cos\theta})+ \mathcal{O}(w^2). \label{Bexp}
\end{equation}
Finally, expanding $\mathrm{e}^{\mathrm{i}\Omega_\pm t/2}$ up to zeroth
order and using (\ref{Bexp}) to compute $c_{10}(t)$ (cf.
Eq.~(\ref{ground_exact})), we get 
\begin{eqnarray}
c_{10}(t) &=&\mathrm{e}^{\mathrm{i} \frac{b t}{2}}\mathrm{e}^{\mathrm{i} \frac{w t}{2}}
\frac{w}{2b}\sin^2\theta\cos\left(\frac{wt}{2}\cos\theta\right)\left[
1 -\mathrm{e}^{-\mathrm{i}bt} \right] \nonumber \\
&&+ \mathcal{O}(w^2). \label{c10}
\end{eqnarray}

\subsubsection{Expansion of $c_{11}(t)$}

Similarly,
\begin{eqnarray}
c_{11}(t) &=&\mathrm{e}^{\mathrm{i} \frac{b t}{2}}\mathrm{e}^{-\mathrm{i} \frac{w t}{2}}
\frac{w}{2b}\sin\theta\!\left[\left( 1 -\mathrm{e}^{-\mathrm{i}bt} \right)
\!\cos\!\left(\!\frac{wt}{2}\cos\theta\!\right)\!\cos\theta 
\right. \nonumber \\
&&\left.-\mathrm{i} \left( 1 +
\mathrm{e}^{-\mathrm{i}bt} \right) \sin\left(\frac{wt}{2}\cos\theta\right) \right]
+ \mathcal{O}(w^2). \label{c11}
\end{eqnarray}

\subsubsection{$|\Psi(t)\rangle$ up to first order in $v=w$}

To first order in $v$, the solution is written as
\begin{equation}
|\Psi(t)\rangle = |\Psi^{(0)}(t)\rangle + v|\Psi^{(1)}(t)\rangle +
\mathcal{O}(v^2).
\end{equation}
Inserting Eqs.~(\ref{c00}), (\ref{c01}), (\ref{c10}), and
(\ref{c11}) into (\ref{cij}) we obtain 
\begin{eqnarray}
\!\!\!\!\!\!\!\!|\Psi^{(0)}(t)\rangle &=&\mathrm{e}^{\mathrm{i} \frac{b t}{2}}\mathrm{e}^{\mathrm{i} \frac{w t}{2}}
\left[ \cos\left(\frac{wt}{2}\cos\theta\right) \right. \label{exact0}\\
&-&\left. \mathrm{i}\cos\theta\sin\left(\frac{wt}{2}\cos\theta\right)\right]|0^0(t)\rangle
\nonumber \\
&+& \mathrm{i}\mathrm{e}^{\mathrm{i} \frac{b t}{2}}\mathrm{e}^{-\mathrm{i} \frac{w
t}{2}}\sin\theta\sin\left(\frac{wt}{2}\cos\theta\right)
|0^1(t)\rangle ,  \nonumber 
\end{eqnarray}
and
\begin{eqnarray}
&|&\hspace{-.3cm}\Psi^{(1)}(t)\rangle = \mathrm{i}\mathrm{e}^{\mathrm{i} \frac{b t}{2}}\mathrm{e}^{\mathrm{i} \frac{w t}{2}}
\frac{w^2t}{4bv}\sin^2\theta\left[
\cos\left(\frac{wt}{2}\cos\theta\right)\right. \nonumber \\
&-&\left. \mathrm{i}\cos\theta\sin\left(\frac{wt}{2}\cos\theta\right)\right]|0^0(t)\rangle
\nonumber \\
&-&\mathrm{e}^{\mathrm{i} \frac{b t}{2}}\mathrm{e}^{-\mathrm{i} \frac{w
t}{2}}\frac{w^2t}{4bv}\sin^3\theta\sin\left(\frac{wt}{2}\cos\theta\right)
|0^1(t)\rangle \nonumber \\
&+&\mathrm{e}^{\mathrm{i} \frac{b t}{2}}\mathrm{e}^{\mathrm{i} \frac{w t}{2}}\frac{w}{2bv}\sin^2\theta
\cos\left(\frac{wt}{2}\cos\theta\right)\left[1-\mathrm{e}^{-\mathrm{i}bt}\right]|1^0(t)\rangle
\nonumber \\
&+&\mathrm{e}^{\mathrm{i} \frac{b t}{2}}\mathrm{e}^{-\mathrm{i} \frac{w t}{2}}\frac{w}{2bv}\sin\theta
\left[\left( 1 -\mathrm{e}^{-\mathrm{i}bt} \right)\cos\theta
\cos\left(\frac{wt}{2}\cos\theta\right) \right.
\nonumber \\
&-&\left.\mathrm{i} \left( 1 +\mathrm{e}^{-\mathrm{i}bt} \right)
\sin\left(\frac{wt}{2}\cos\theta\right) \right]|1^1(t)\rangle.
\label{exact1}
\end{eqnarray}

\subsection{Comparison with DAPT}

\subsubsection{Computing the Wilczek-Zee phase}

To determine the zeroth and first order contributions in DAPT, we have to 
perform some
previous calculations. We need to explicitly compute
the non-abelian geometric WZ-phase since it appears in the corrections coming
from DAPT.  In particular, we must solve explicitly 
\begin{equation}
\dot{\mathbf{U}}^n(s) + \mathbf{U}^{n}(s)\mathbf{M}^{nn}(s)=0,
\label{difWZs}
\end{equation}
whose formal solution is
\begin{equation}
\mathbf{U}^{n}(s) = \mathbf{U}^{n}(0)\mathcal{T}
\exp\left( -\int_0^s\mathbf{M}^{nn}(s')ds'\right).
\end{equation}
In general, the solution to the coupled differential equations coming from 
(\ref{difWZs}) cannot be put into a closed form. 
Fortunately, for the model we are dealing with such exact closed solution exists.

Since our model has two doubly degenerate eigenvalues, Eq.~(\ref{difWZs}), 
$\dot{U}^n_{h_ng_n}(s) + \sum_{k_n=0}^{1}U^n_{h_nk_n}(s) M^{nn}_{g_nk_n}(s)=0$, 
reduces to two sets ($n=0,1$) of four equations ($h_ng_n=00,01,10,11$):
\begin{eqnarray} \dot{U}^n_{00}(s) + U^n_{00}(s)
M^{nn}_{00}(s)+U^n_{01}(s)
M^{nn}_{01}(s)&=&0, \label{U00}\\
\dot{U}^n_{01}(s) + U^n_{00}(s) M^{nn}_{10}(s)+U^n_{01}(s)
M^{nn}_{11}(s)&=&0, \label{U01}\\
\dot{U}^n_{10}(s) + U^n_{10}(s) M^{nn}_{00}(s)+U^n_{11}(s)
M^{nn}_{01}(s)&=&0, \label{U10}\\
\dot{U}^n_{11}(s) + U^n_{10}(s) M^{nn}_{10}(s)+U^n_{11}(s)
M^{nn}_{11}(s)&=&0. \label{U11}
\end{eqnarray}
Note that the first two equations are not coupled to the last
two, which is one of the ingredients for solvability. 

Let us start computing the four matrices below
(cf. Eq.~(\ref{M})),
\begin{equation}
\mathbf{M}^{mn}(s)=  \left(
\begin{array}{cccc}
M^{nm}_{00}(s)& M^{nm}_{10}(s) \\
M^{nm}_{01}(s) & M^{nm}_{11}(s)
\end{array}
\right),
\end{equation}
with $n,m=0,1$ and $M^{nm}_{h_ng_m}(s)=\langle
n^{h_n}(s)|\dot{m}^{g_m}(s) \rangle$.
An easy calculation using
Eqs.~(\ref{zerozerostate})-(\ref{oneonestate}) leads to
\begin{equation}
\mathbf{M}^{nm}(s) = 
%
%
\left(
\begin{array}{cccc}
-\frac{\mathrm{i}w}{2v}\sin^2\theta & -\frac{\mathrm{i}w}{4v}\sin(2\theta)\mathrm{e}^{-\mathrm{i}\frac{ws}{v}} \\
-\frac{\mathrm{i}w}{4v}\sin(2\theta)\mathrm{e}^{\mathrm{i}\frac{ws}{v}} &
\frac{\mathrm{i}w}{2v}\sin^2\theta
\end{array}
\right). \label{Mnm}
\end{equation}
Therefore, since
$\mathbf{M}^{00}(s)=\mathbf{M}^{11}(s)$, then  
$\mathbf{U}^{0}(s)=\mathbf{U}^{1}(s)$, and we have to solve
two pairs of coupled equations. Moreover, the first pair of
equations, (\ref{U00}) and (\ref{U01}), are formally equivalent to
the last one, (\ref{U10}) and (\ref{U11}). 

Using Eq.~(\ref{Mnm})
they can be written as follows,
\begin{eqnarray} \dot{x}(s) -a x(s)
+b\mathrm{e}^{\mathrm{i}cs}y(s) &=&0, \label{x}\\
\dot{y}(s) +a y(s) +b\mathrm{e}^{-\mathrm{i}cs}x(s) &=&0, \label{y}
\end{eqnarray}
with
\begin{eqnarray*}
a = \frac{\mathrm{i}w}{2v}\sin^2\theta,\hspace{.3cm} & b = -\frac{\mathrm{i}w}{4v}\sin(2\theta),
\hspace{.3cm}& c=\frac{w}{v},
\end{eqnarray*}
and $(x(s),y(s))=(U^n_{00}(s),U^n_{01}(s))$ or
$(x(s),y(s))=(U^n_{10}(s),U^n_{11}(s))$. There is only one subtle
difference between the equations giving either
$(U^n_{00}(s),U^n_{01}(s))$ or $(U^n_{10}(s),U^n_{11}(s))$. It is
the initial condition. Since we adopted $\mathbf{U}^{n}(0)=\mathds{1}$, 
$\mathds{1}$
being the identity, we have either
$(U^n_{00}(0),U^n_{01}(0))=(1,0)$ or
$(U^n_{10}(0),U^n_{11}(0))=(0,1)$. 

In order to solve these coupled
differential equations we make the following change of variables
$$\tilde{x}(s)=\mathrm{e}^{-\mathrm{i}cs/2}x(s) \hspace{.5cm} \mbox{and}
\hspace{.5cm} \tilde{y}(s)=\mathrm{e}^{\mathrm{i}cs/2}y(s),$$
which leads to
\begin{eqnarray} \dot{\tilde{x}}(s) + (\mathrm{i}c/2 - a) \tilde{x}(s)
+b \tilde{y}(s) &=&0, \label{xtilde}\\
\dot{\tilde{y}}(s) +(a-\mathrm{i}c/2) \tilde{y}(s) +b \tilde{x}(s) &=&0.
\label{ytilde}
\end{eqnarray}
Now we have two coupled linear first order differential equations with constant
coefficients that can be easily decoupled and solved in closed form.
Hence, solving the equations above and returning to the original
variables we finally get the WZ-phase,
\begin{eqnarray}
\mathbf{U}^{n}(s) &=& \left(
\begin{array}{cccc}
U^{n}_{00}(s)& U^{n}_{01}(s) \\
U^{n}_{10}(s) & U^{n}_{11}(s)
\end{array}
\right)
\label{UWZ},
\end{eqnarray}
where for $n=0$ or $1$ we obtain
\begin{eqnarray*}
U^{n}_{00}(s) & = & [U^{n}_{11}(s)]^* \\
&=&\mathrm{e}^{\mathrm{i}\frac{ws}{2v}}\!\!\left[ \cos\left(\frac{ws}{2v}\cos\theta\right)\right.
\left. -\mathrm{i} \cos\theta\sin\left(\frac{ws}{2v}\cos\theta\right) \right]\!,
\\
U^{n}_{01}(s) & = & 
-[U^{n}_{10}(s)]^*=\mathrm{i}\mathrm{e}^{-\mathrm{i}\frac{ws}{2v}}\sin\theta \sin\left(\frac{ws}{2v}\cos\theta\right).
\end{eqnarray*}

\subsubsection{Zeroth order correction}

For our example, Eq.~(\ref{zeroth}) becomes
\begin{eqnarray*}
|\Psi^{(0)}(s)\rangle\! &=&
\mathrm{e}^{-\frac{\mathrm{i}}{v}\omega_{0}(s)}
U^{0}_{00}(s)|0^{0}(s)\rangle
\!+\!\mathrm{e}^{-\frac{\mathrm{i}}{v}\omega_{0}(s)}
U^{0}_{01}(s)|0^{1}(s)\rangle.
\end{eqnarray*}
Using Eq.~(\ref{UWZ}) and remembering that
$\omega_0(s)=-bs/2=-bvt/2$ and $s=vt$ we get,
\begin{eqnarray}
|\Psi^{(0)}(t)\rangle &=&\mathrm{e}^{\mathrm{i}\frac{bt}{2}}\mathrm{e}^{\mathrm{i}\frac{wt}{2}}
\left[ \cos\left(\frac{wt}{2}\cos\theta\right)\right. \nonumber \\
&-&\left.  \mathrm{i}
\cos\theta\sin\left(\frac{wt}{2}\cos\theta\right)\right]
|0^{0}(t)\rangle \nonumber \\
\hspace{-.15cm}&+&\hspace{-.15cm}\mathrm{i}\mathrm{e}^{\mathrm{i}\frac{bt}{2}}\!\mathrm{e}^{-\mathrm{i}\frac{wt}{2}}\!\sin\theta
\sin\!\left(\!\frac{wt}{2}\cos\theta\!\right)\!\! |0^{1}(t)\rangle\!,
\label{exampleAd}
\end{eqnarray}
which is identical to the expression obtained by expanding the exact solution up to
zeroth order (Eq.~(\ref{exact0})).

\subsubsection{First order correction}

Since we have only two doubly degenerate eigenvalues 
Eq.~(\ref{first}) becomes
\begin{eqnarray}
&|&\hspace{-.3cm}\Psi^{(1)}(s)\rangle=
\mathrm{i}\hbar \mathrm{e}^{-\frac{\mathrm{i}}{v}\omega_0(s)}
[\mathbf{J}^{010}(s) \mathbf{U}^0(s)]_{00}|0^{0}(s)\rangle
\nonumber \\
&+&\mathrm{i}\hbar \mathrm{e}^{-\frac{\mathrm{i}}{v}\omega_0(s)}
[\mathbf{J}^{010}(s) \mathbf{U}^0(s)]_{01}|0^{1}(s)\rangle
\nonumber \\
&-&\mathrm{i}\hbar\mathrm{e}^{-\frac{\mathrm{i}}{v}\omega_1(s)}
\frac{[\mathbf{U}^0(0)\mathbf{M}^{01}(0)
\left(\mathbf{U}^1(0)\right)^\dagger
\mathbf{U}^1(s)]_{00}}{\Delta_{10}(0)}|1^{0}(s)\rangle
\nonumber \\
&-&\mathrm{i}\hbar \mathrm{e}^{-\frac{\mathrm{i}}{v}\omega_1(s)}
\frac{[\mathbf{U}^0(0)\mathbf{M}^{01}(0)
\left(\mathbf{U}^1(0)\right)^\dagger
\mathbf{U}^1(s)]_{01}}{\Delta_{10}(0)}|1^{1}(s)\rangle
\nonumber \\
&+&\mathrm{i}\hbar\mathrm{e}^{-\frac{\mathrm{i}}{v}\omega_0(s)}
\frac{[\mathbf{U}^0(s)\mathbf{M}^{01}(s)]_{00}}
{\Delta_{10}(s)}|1^{0}(s)\rangle\nonumber \\
&+&\mathrm{i}\hbar\mathrm{e}^{-\frac{\mathrm{i}}{v}\omega_0(s)}
\frac{[\mathbf{U}^0(s)\mathbf{M}^{01}(s)]_{01}}
{\Delta_{10}(s)}|1^{1}(s)\rangle.
\end{eqnarray}

We can re-write it as follows by noting
that our initial conditions imply
that $\mathbf{U}^n(0)=\mathds{1}$,
\begin{eqnarray}
|\Psi^{(1)}(s)\rangle&=&
\mathrm{i}\hbar \mathrm{e}^{-\frac{\mathrm{i}}{v}\omega_0(s)}
[\mathbf{J}^{010}(s) \mathbf{U}^0(s)]_{00}|0^{0}(s)\rangle
\nonumber \\
&+&\mathrm{i}\hbar \mathrm{e}^{-\frac{\mathrm{i}}{v}\omega_0(s)}
[\mathbf{J}^{010}(s) \mathbf{U}^0(s)]_{01}|0^{1}(s)\rangle
\nonumber \\
&-&\mathrm{i}\hbar\left(\mathrm{e}^{-\frac{\mathrm{i}}{v}\omega_1(s)}
\frac{[\mathbf{M}^{01}(0)\mathbf{U}^1(s)]_{00}}{\Delta_{10}(0)}\right.
\nonumber \\
&-&\left.\mathrm{e}^{-\frac{\mathrm{i}}{v}\omega_0(s)}
\frac{[\mathbf{U}^0(s)\mathbf{M}^{01}(s)]_{00}}
{\Delta_{10}(s)}\right)|1^{0}(s)\rangle
\nonumber \\
&-&\mathrm{i}\hbar
\left(\mathrm{e}^{-\frac{\mathrm{i}}{v}\omega_1(s)}
\frac{[\mathbf{M}^{01}(0)\mathbf{U}^1(s)]_{01}}{\Delta_{10}(0)} 
\right.\nonumber \\
\hspace{-.2cm}&-&\hspace{-.2cm}\left.\mathrm{e}^{-\frac{\mathrm{i}}{v}\omega_0(s)}
\frac{[\mathbf{U}^0(s)\mathbf{M}^{01}(s)]_{01}} {\Delta_{10}(s)}
\right)|1^{1}(s)\rangle,
\end{eqnarray}
with
\begin{equation}
\mathbf{J}^{010}(s) =\int_0^s\mathrm{d}s'
\left(\frac{\mathbf{U}^0(s')\mathbf{M}^{01}(s')\mathbf{M}^{10}(s')
\left(\mathbf{U}^0(s')\right)^\dagger}{\Delta_{01}(s')}\right).
\label{J010}
\end{equation}

A direct computation gives
\begin{eqnarray}
\mathbf{J}^{010}(s) &=& \left(
\begin{array}{cccc}
\frac{w^2s\sin^2\theta}{4v^2\hbar b} & 0 \\
0 & \frac{w^2s\sin^2\theta}{4v^2\hbar b}
\end{array}
\right).
\end{eqnarray}
%
%
%
%
%
%

Putting all these pieces together 
we get
\begin{eqnarray}
&|\hspace{-.3cm}&\Psi^{(1)}(t)\rangle = \mathrm{i}\mathrm{e}^{\mathrm{i} \frac{b t}{2}}\mathrm{e}^{\mathrm{i} \frac{w t}{2}}
\frac{w^2t}{4bv}\sin^2\theta\left[
\cos\left(\frac{wt}{2}\cos\theta\right) \right.\nonumber \\
&&\left. -\mathrm{i}\cos\theta\sin\left(\frac{wt}{2}\cos\theta\right)\right]|0^0(t)\rangle
\nonumber \\
&&-\mathrm{e}^{\mathrm{i} \frac{b t}{2}}\mathrm{e}^{-\mathrm{i} \frac{w
t}{2}}\frac{w^2t}{4bv}\sin^3\theta\sin\left(\frac{wt}{2}\cos\theta\right)
|0^1(t)\rangle \nonumber \\
&&+\mathrm{e}^{\mathrm{i} \frac{b t}{2}}\mathrm{e}^{\mathrm{i} \frac{w t}{2}}\frac{w}{2bv}\sin^2\theta
\cos\left(\frac{wt}{2}\cos\theta\right)\left[1-\mathrm{e}^{-\mathrm{i}bt}\right]|1^0(t)\rangle
\nonumber \\
&&+\mathrm{e}^{\mathrm{i} \frac{b t}{2}}\mathrm{e}^{-\mathrm{i} \frac{w t}{2}}\frac{w}{2bv}\sin\theta
\left[\left( 1 -\mathrm{e}^{-\mathrm{i}bt} \right)\cos\theta
\cos\left(\frac{wt}{2}\cos\theta\right) \right.
\nonumber \\
&&\left.-\mathrm{i} \left( 1 +\mathrm{e}^{-\mathrm{i}bt} \right)
\sin\left(\frac{wt}{2}\cos\theta\right) \right]|1^1(t)\rangle,
\label{firstA}
\end{eqnarray}
which is exactly Eq.~(\ref{exact1}), the first order correction
obtained from expanding the exact solution.

It is worth noting that the first order correction terms associated 
to the degenerate eigenspace $\mathcal{H}^0$, i.e., 
the first two terms of Eq.~(\ref{firstA}), do not appear in standard
approaches trying to correct DAA. In general we only see first order
terms related to the excited eigenspaces. This same feature is
seen for the non-degenerate case \cite{Rig08}, where the corresponding term
is also missing \cite{Tho83,Che11}. However,
as the expansion of the exact solution clearly demonstrates, 
these terms must appear (and they do for DAPT) 
in any perturbation theory about DAA.

\subsection{The necessary condition}

The necessary condition (\ref{strongerMat}) for our example,
where we only have two eigenspaces ($n=0,1$), looks like
\begin{equation}
\hbar\left\|\frac{\mathbf{M}^{10}(t)}{\Delta_{10}(t)}\right\|_1
\ll 1,  \hspace{.5cm} t \in [0,T].
\end{equation}

Using Eq. (\ref{Mnm}) the necessary condition reads
\begin{equation}
\frac{w}{2b}\left|\sin^2\theta + \frac{\sin(2\theta)}{2}\right| \ll 1
\Longrightarrow \frac{w\sin\theta}{2b}|\sin\theta + \cos\theta| \ll 1.
\label{n1}
\end{equation}
Note that since the maximum of $|\sin\theta + \cos\theta|$ is $\sqrt{2}$
the condition $\frac{w\sin\theta}{b}\ll1$ is stronger, i.e., it implies
(\ref{n1}). Also, since $\sin\theta$
cannot exceed $1$, $\frac{w}{b}\ll1$ is even stronger.
Our task at this moment is to look at the exact solution, 
assume that DAA holds, and prove it implies one of the necessary conditions above. 

If DAA
holds then the absolute values of the
coefficients multiplying $|1^0(t)\rangle$ and $|1^1(t)\rangle$ must
be negligible. Therefore, looking at Eq. (\ref{ground_exact}) we must have
\begin{eqnarray}
\frac{\sin^2\theta}{2} \left| B_+(t) +
B_-(t)\right| \ll 1, 
\end{eqnarray}
\begin{eqnarray}
\hspace{-.5cm}\frac{1}{2} \sin\theta \left| (1 + \cos\theta)
B_-(t) - (1 - \cos\theta) B_+(t) \right| \ll 1.
\end{eqnarray}
Now, using Eq. (\ref{Bpm}) it implies
\begin{eqnarray}
\frac{w\sin^2\theta}{2} \left| \frac{\sin(\Omega_+ t /2)}{\Omega_+} +
\frac{\sin(\Omega_- t /2)}{\Omega_-}\right| \ll 1,
\end{eqnarray}
\begin{eqnarray}
\frac{w\!\sin\!\theta}{2}\!  \left|\! (1 \!+\! \cos\!\theta)
\frac{\sin(\Omega_- t /2)}{\Omega_-} \!-\!
(1 \!-\! \cos\!\theta) \frac{\sin(\Omega_+ t /2)}{\Omega_+}\! \right|\! \ll\! 1.
\nonumber \\
\end{eqnarray}
Since we want these coefficients to be very small at all times, let us
work with the worst scenario for each one of the inequalities, i.e.,
when $\sin(\Omega_\pm t /2)\approx 1$ for the first one and
$\sin(\Omega_\pm t /2)\approx \mp1$ for the second one. This last worst case scenario
occurs since $0\leq \theta\leq \pi$ implies $1\pm\cos\theta\geq0$ and
we need a minus sign coming from $\sin(\Omega_+ t /2)$ to compensate the minus sign
before $1-\cos\theta$; all quantities must be positive  in the worst case
scenario.
Hence, those conditions become
\begin{eqnarray}
\frac{w\sin^2\theta}{2} \left| \frac{1}{\Omega_+} +
\frac{1}{\Omega_-}\right| &\ll& 1, \\
\frac{w\sin\theta}{2}  \left| (1 + \cos\theta)
\frac{1}{\Omega_-} +
(1 - \cos\theta) \frac{1}{\Omega_+} \right| &\ll& 1.
\end{eqnarray}
Re-writing the last inequality we obtain
\begin{eqnarray}
\frac{w\sin^2\theta}{2} \left| \frac{1}{\Omega_+} +
\frac{1}{\Omega_-}\right| \hspace{-.1cm}&\ll&\hspace{-.1cm} 1, \\
\frac{w\sin\theta}{2}  \left|\frac{1}{\Omega_+} +
\frac{1}{\Omega_-} + \cos\theta
\left(\frac{1}{\Omega_-} - \frac{1}{\Omega_+}\right) \right| \hspace{-.1cm}&\ll&\hspace{-.1cm} 1.
\end{eqnarray}

Actually, these two inequalities are not independent. If the second one is
satisfied, so is the first one. To see this note that we can write the first one as
\begin{eqnarray}
\frac{w\sin\theta}{2} \left| \sin\theta\left(\frac{1}{\Omega_+} +
\frac{1}{\Omega_-}\right)\right| &\ll& 1.
\end{eqnarray}
Both inequalities have the same factor multiplying the moduli, $(w\sin\theta)/2$.
Hence, if we are able to show that the second absolute value is always greater than
the first one, we prove that the second inequality implies the first one. First, we
note that we always have $\Omega_\pm\geq 0$. Then, looking at the term
$\sin\theta\left(1/\Omega_+ + 1/\Omega_-\right)$ of the first
inequality and $1/\Omega_+ + 1/\Omega_-$ of the second, we realize
that the latter is always greater than the former since $|\sin\theta|\leq 1$. This
means that if we show that the other term of the second inequality, $\cos\theta
\left(1/\Omega_- - 1/\Omega_+\right)$, is always
positive, we prove our claim. This proof is divided in two parts. We first analyze
the case where $\theta \in [0,\pi/2]$ and then the case $\theta \in (\pi/2,\pi]$.
These two intervals cover the whole span of the polar angle $\theta$.

Remembering that $\Omega^2_\pm$ $=$ $w^2 + b^2 \pm 2wb\cos\theta$,
Eq. (\ref{Omega}), we readily see that for  $\theta \in [0,\pi/2]$
we must have $\Omega_+ \geq \Omega_-$. This gives
$1/\Omega_- - 1/\Omega_+\geq 0$. But in this interval
$\cos\theta\geq0$ implying that
$\cos\theta\left(1/\Omega_- - 1/\Omega_+\right)\geq0$.
For  $\theta \in (\pi/2,\pi]$, on the other hand, $\Omega_+ < \Omega_-$,
which in turn leads to $1/\Omega_- - 1/\Omega_+ < 0$. Since now
$\cos\theta<0$ then $\cos\theta\left(1/\Omega_- - 1/\Omega_+\right)\geq0$
too, completing our proof.

In other words, we just need to focus on the following inequality
\begin{eqnarray}
\frac{w\sin\theta}{2b}  f(\theta) &\ll& 1,
\label{aac2}
\end{eqnarray}
with
\begin{equation}
f(\theta)=\left|\frac{b}{\Omega_+} +
\frac{b}{\Omega_-} + \cos\theta
\left(\frac{b}{\Omega_-} - \frac{b}{\Omega_+}\right) \right|,
\label{aac3}
\end{equation}
which follows from assuming DAA holds and
our task is to show that (\ref{aac2}) implies the 
necessary condition (\ref{n1}).

Note that for any $w,b>0$ (assumed when we
solved Hamiltonian (\ref{Hb}))
and $\theta \in [0,\pi]$ (polar angle)
Eq. (\ref{aac3}) has a global minimum at $\theta=\pi/2$. 
Therefore, $f(\theta)\geq f(\pi/2)=2b/\sqrt{b^2+w^2}$. When
$w<b$ we have,
\begin{eqnarray*}
f(\theta)\geq \frac{2b}{\sqrt{b^2+w^2}} =\frac{2}{\sqrt{1+w^2/b^2}}
\geq\frac{2}{\sqrt{1+1}}=\sqrt{2}.
\end{eqnarray*}
The last inequality results from the fact that $w<b$.
With this lower bound the lhs of
(\ref{aac2}) becomes
\begin{eqnarray}
\hspace{-.5cm}\frac{w\sin\theta}{2b}  f(\theta) \geq \frac{w\sin\theta}{2b}  \sqrt{2}
\geq \frac{w\sin\theta}{2b}  |\sin\theta + \cos\theta|,
\end{eqnarray}
where the last inequality is a consequence of $\sqrt{2}$ being the maximum
of $|\sin\theta + \cos\theta|$. Equivalently,
\begin{equation}
\frac{w\sin\theta}{2b}  |\sin\theta + \cos\theta| \leq
\frac{w\sin\theta}{2b}  f(\theta).
\end{equation}
But the lhs above is just the expression coming from
the necessary condition. Hence, since whenever the adiabatic
approximation holds $\frac{w\sin\theta}{2b}  f(\theta) \ll 1$, we have that
the necessary condition is automatically satisfied.

For completeness, let us analyze what happens for $w\geq b$, when it is expected that DAA does not hold 
since the rotating frequency $w$ of the magnetic field is greater or equal to $b$, i.e., the 
Hamiltonian changes in a rate ($w$) 
at least as big as the internal characteristic frequency of the system ($b$). 
Note also that unless $\theta\approx 0$ the necessary condition
cannot be satisfied either (cf. Eq. (\ref{n1})). 

In this case
\begin{eqnarray*}
f(\theta)\geq \frac{2b}{\sqrt{b^2+w^2}} =\frac{2b}{w}\frac{1}{\sqrt{1+b^2/w^2}}
\geq\frac{b}{w}\sqrt{2}.
\end{eqnarray*}
This implies that the lhs of Eq. (\ref{aac2}) becomes
\begin{eqnarray}
\frac{w\sin\theta}{2b}  f(\theta) > \frac{w\sin\theta}{2b} \frac{b}{w}\sqrt{2}
= \sin\theta \frac{\sqrt{2}}{2} \approx \sin\theta.
\label{aux1}
\end{eqnarray}
For not too small $\theta$ we have $\sin\theta \approx 1$ and it is clear that the system is not described by the adiabatic approximation. Indeed, Eq.~(\ref{aux1})
implies that at least one of the coefficients multiplying the excited states $|1^0(t)\rangle$ or $|1^1(t)\rangle$ is of order $\sin\theta$.  

When in addition to $w\geq b$ we have $\theta \approx 0$,
it can be shown that DAA continues to be a bad approximation to the evolution of the system, 
despite the fact that the fidelity between the exact solution and DAA approaches one and that 
the necessary condition is satisfied. 
As shown in Appendix \ref{appendixD}, if $w\geq b$ and $\theta \approx 0$ 
the probability to measure
the system at the excited state $|1^1(t)\rangle$ is of the same order in $\theta$ as that 
of measuring it in $|0^1(t)\rangle$. 
This clearly indicates that the necessary condition is not a sufficient one.  
However, the sufficient condition derived in the next section excludes this case
as an instance where one can approximate the system's evolution by DAA.

\subsection{The sufficient condition}

For the specific problem we are dealing with 
the sufficient conditions,
Eqs. (\ref{suf1at}) and (\ref{suf1bt}), become
\begin{widetext}
\begin{equation}
2\hbar \int_0^t\mathrm{d}t'
\left(\frac{\sum_{k_0,i_0=0}^{1}|[\mathbf{M}^{01}(t')
(\mathbf{M}^{01}(t'))^\dagger]_{k_0i_0}|}{|\Delta_{01}(t')|}\right)
\ll \min \limits_{\forall g_0}\!_+
\left(\left|\left[\mathbf{U}^{0}(t)\right]_{0g_0}\right|\right),
\hspace{.5cm} t \in [0,T],
\label{suf1a_ex}
\end{equation}
and
\begin{eqnarray}
\frac{\hbar}{|\Delta_{10}(0)|}\left(\sum_{k_0=0}^{1}
\hspace{-.15cm}\left|[\mathbf{M}^{01}(t)]_{k_0g_1}\right| +
2\sum_{k_0,l_1=0}^{1}
\left| [\mathbf{M}^{01}(0)]_{k_0l_1} \right|\right)
&\ll&\min \limits_{\forall g_0}\!_+
\left(\left|\left[\mathbf{U}^{0}(t)\right]_{0g_0}\right|\right),\;
\forall g_1, \; t \in [0,T]. \label{suf1b_ex}
\end{eqnarray}
\end{widetext}
Note that this last equation encompasses two instances,
$g_1=0$ and $g_1=1$. Let us simplify each one separately.

Using (\ref{Mnm}) the numerator of Eq. (\ref{suf1a_ex}) can
be written as
\begin{equation}
\sum_{k_0,i_0=0}^{1}|[\mathbf{M}^{01}(t)
(\mathbf{M}^{01}(t))^\dagger]_{k_0i_0}| = \frac{w^2\sin^2\theta}{2}.
\end{equation}
Hence, Eq.~(\ref{suf1a_ex}) is simply
\begin{equation}
\frac{w^2t}{b}\sin^2\theta\ll \min \limits_{\forall g_0}\!_+
\left(\left|\left[\mathbf{U}^{0}(t)\right]_{0g_0}\right|\right),
\hspace{.5cm} t \in [0,T].
\label{s1}
\end{equation}

Moving our attention to the other sufficient condition we first note
that $\sum_{k_0=0}^{1}
\left|[\mathbf{M}^{01}(t)]_{k_0g_1}\right|$ gives the same sum whether
$g_1=0$ or $g_1=1$. In other words, we have only one case to consider. 
Using Eq. (\ref{Mnm}) a direct calculation gives
\begin{eqnarray*}
\frac{5w}{4b}(|\sin(2\theta)|+2\sin^2\theta)
&\!\ll\!&\min \limits_{\forall g_0}\!_+
\left(\left|\left[\mathbf{U}^{0}(t)\right]_{0g_0}\right|\right),
\; t \in [0,T]
\end{eqnarray*}

But since $\theta \in [0,\pi]$ we can write the previous expression as

\begin{eqnarray}
\frac{5w}{2b}\sin\theta(|\cos\theta|+\sin\theta)
&\!\ll\!&\min \limits_{\forall g_0}\!_+
\left(\left|\left[\mathbf{U}^{0}(t)\right]_{0g_0}\right|\right),
\; t \in [0,T].\nonumber \\
\label{s2}
\end{eqnarray}

Using the WZ-phase (\ref{UWZ}) it is not
difficult to see that
$$\left|\left[\mathbf{U}^{0}(t)\right]_{00}\right|=\sqrt{1-\sin^2\theta
\sin^2\left(\frac{wt\cos\theta}{2}\right)}$$ and
$$\left|\left[\mathbf{U}^{0}(t)\right]_{01}\right|=\sin\theta
\left|\sin\left(\frac{wt\cos\theta}{2}\right)\right|.$$

Furthermore, in this example we chose
$v=w=1/T$ which implies that $wt\leq 1$ during the whole evolution.
Thus $|\sin(wt\cos(\theta)/2)|\leq \sin(1/2)$ leading to
$$|[\mathbf{U}^{0}(t)]_{00}|
\geq |[\mathbf{U}^{0}(t)]_{01}|$$ and to 
$$\frac{5w}{2b}\sin\theta(|\cos\theta|+\sin\theta)\geq 
\frac{w^2t}{b}\sin^2\theta.$$
With these two inequalities we see that Eqs.~(\ref{s1}) and (\ref{s2})
collapse to the following sufficient condition,
\begin{eqnarray}
\frac{5w}{2b}
&\!\ll\!&
\frac{\left|\sin\left(\frac{wt\cos\theta}{2}\right)\right|}
{|\cos\theta|+\sin\theta},
\;\; t \in [0,T].
\label{s1e2}
\end{eqnarray}
Notice that for $t\approx 0$ and $\theta\approx \pi/2$ we have 
$|[\mathbf{U}^{0}(t)]_{01}|\approx 0$ and we need to work 
with the non-null coefficient $|[\mathbf{U}^{0}(t)]_{00}|$.
In this scenario the sufficient condition becomes $5w/(2b)\ll1$.

First thing we note is that, at least for this example, the sufficient
condition is stronger than, and implies, the necessary
condition. To see this, take Eq. (\ref{s2}). It is not
difficult to see that
\begin{displaymath}
\frac{5w}{2b}\sin\theta(|\cos\theta|+\sin\theta) \geq
\frac{w}{2b}\sin\theta|\cos\theta+\sin\theta|.
\end{displaymath}
Hence, if Eq. (\ref{s2}) is satisfied we automatically have
\begin{eqnarray*}
\frac{w}{2b}\sin\theta|\cos\theta+\sin\theta| \ll 1,
\end{eqnarray*}
which is exactly the necessary condition, Eq. (\ref{n1}).
See also Ref. \cite{Yuk09} for an alternative route to establish sufficient conditions
in non-degenerate systems, claimed to be general and in some cases also necessary.

Second, if $w\geq b$ we cannot satisfy the sufficient condition
(\ref{s1e2}) irrespective of the value of $\sin\theta$. This is true because
the rhs of (\ref{s1e2}) is never greater than one. Thus, if
$w\geq b$ the lhs is always greater than one and the inequality
cannot be satisfied at all. This is a very satisfactory restriction that
shows the sufficient condition is consistent with the cases where
the necessary one fails. 

To complete the analysis we just need to show that for $w<b$ the sufficient
condition implies DAA. In other words, we must show that 
the absolute values of the coefficients multiplying $|1^0(t)\rangle$ and $|1^1(t)\rangle$
are negligible if the sufficient condition holds.


To show that in a clear and straightforward manner we first need
to manipulate algebraically  those coefficients. Let us call them
$C_{|1^0(t)\rangle}$ and $C_{|1^1(t)\rangle}$. From Eq.
(\ref{ground_exact}) and remembering that $\theta\in [0,\pi]$ 
we have
\begin{eqnarray*}
|C_{|1^0(t)\rangle}| &\!=\!& \frac{\sin^2\theta}{2}\! \left| B_+(t) \!+\!
B_-(t)\right|\!\leq\! \frac{\sin\theta}{2}\! \left| B_+(t) +
B_-(t)\right|\!,\\
|C_{|1^1(t)\rangle}| &=& \frac{\sin\theta}{2} \left| (1 +
\cos\theta) B_-(t) - (1 - \cos\theta) B_+(t) \right|\nonumber
\\
&\leq& \frac{\sin\theta}{2} \left| (1 + \cos\theta) B_-(t) + (1 -
\cos\theta) B_+(t) \right|  \nonumber \\
&=& \frac{\sin\theta}{2} \left| B_+(t) + B_-(t) +\cos\theta
(B_-(t) - B_+(t)) \right|.
\end{eqnarray*}
Using Eq. (\ref{Bpm}) and the maximum value possible for 
$\sin(\Omega_\pm t /2)$ we have
\begin{eqnarray}
\hspace{-.5cm}|C_{|1^0(t)\rangle}| &\!\leq\!&  \frac{w\!\sin\theta}{2b}\! \left|
\frac{b}{\Omega_+} \!+\! \frac{b}{\Omega_-}\right|,\label{coef10}\\
\hspace{-.5cm}|C_{|1^1(t)\rangle}| &\!\leq\!& \frac{w\!\sin\theta}{2b}\! \left|\!
\frac{b}{\Omega_+} \!+\! \frac{b}{\Omega_-} \!+\! |\cos\theta|\!
\left(\!\frac{b}{\Omega_-} \!+\! \frac{b}{\Omega_+}\!\right)\! \right|\!
\label{coef11}.
\end{eqnarray}
It is obvious that the rhs of the last
inequality is greater than the rhs of the first one.
Hence, if we
show that the sufficient conditions imply that the rhs
of Eq. (\ref{coef11}) is much smaller than one the proof is accomplished.

To this end we write the rhs of Eq. (\ref{coef11}) as follows
\begin{eqnarray}
{\sf rhs}(\theta) &=& \frac{w}{2b}(1 + |\cos\theta|)
\sin\theta\left(\frac{b}{\Omega_+} + \frac{b}{\Omega_-}\right)
\nonumber \\
&\leq& \frac{w}{b}
\sin\theta\left(\frac{b}{\Omega_+} + \frac{b}{\Omega_-}\right).
\end{eqnarray}
But one can show that the function
$$
g(\theta)=\sin\theta\left(\frac{b}{\Omega_+}
+ \frac{b}{\Omega_-}\right)
$$
has a maximum for $\theta \in [0,\pi]$ at $\theta=\pi/2$ given by
$g_{\sf max}=2b/\sqrt{b^2+w^2}.$
Therefore,
\begin{eqnarray*}
{\sf rhs}(\theta) &\leq& \frac{w}{b}g_{\sf max}
= \frac{2w}{\sqrt{b^2+w^2}} = \frac{2w}{b}\frac{1}{\sqrt{1+w^2/b^2}}
\leq \frac{2w}{b}. 
\end{eqnarray*}

From (\ref{s1e2}) we can show for $t\in [0,T]$ that
$$
\frac{\left|\sin\left(\frac{wt\cos\theta}{2}\right)\right|}
{|\cos\theta|+\sin\theta} \leq \sin(1/2)\approx 0.48<1/2,
$$ 
since $wt\leq 1$. Therefore, the sufficient condition (\ref{s1e2})
and the case where $\theta\approx \pi/2$ reduces to
\begin{equation}
5w/b \ll 1, \;\; t \in [0,T],
\end{equation}
which obviously implies 
$
2w/b\ll 1,
$
the condition needed to have all coefficients of the excited 
eigenspace negligible.

\section{Numerical examples}
\label{examples}

We now want to test DAPT for other degenerate Hamiltonians with and
without a constant gap. For that purpose we work with the following Hamiltonian,
already written in the rescaled time $s$,
\begin{equation}
\mathbf{H}(s) =\frac{1}{\sqrt{2}}
\left( 
\begin{array}{cc}
\mathbf{0}&\mathbf{H}_1(s)\\
\mathbf{H}_1^\dagger(s)&\mathbf{0}
\end{array}
\right),
\label{otherH}
\end{equation}
where
\begin{eqnarray}
\mathbf{H}_1(s) &=&
\left( 
\begin{array}{cccc}
-E(s)&\mathrm{e}^{-\mathrm{i}\theta(s)}E(s)\\
\mathrm{e}^{\mathrm{i}\theta(s)}E(s)&E(s)
\end{array}
\right),\\
E(s) &=& E_0 + \lambda (s-1/2)^2,
\label{lamb}\\
\theta(s) &=& \theta_0 + ws^2.
\end{eqnarray}
Note that for $\lambda=0$ the gap is constant while for $\lambda > 0$
it changes quadratically in time achieving its minimum value at $s=1/2$. 

Hamiltonian (\ref{otherH}) is a doubly degenerate system with eigenvalues
given by $-E(s)$ and $E(s)$, and corresponding eigenvectors
\begin{eqnarray}
|0^0(s)\rangle\! &=&\! \frac{1}{2}\!\left(\mathrm{e}^{-\mathrm{i}\theta(s)}|\!\!\uparrow \uparrow \rangle
+|\!\!\uparrow\downarrow\rangle -\sqrt{2}|\!\!\downarrow\downarrow\rangle\!\right)\!\!,
\label{newzerozerostate}\\
|0^1(s)\rangle\! &=&\! \frac{1}{2}\!\left(\!| \!\!\uparrow \uparrow \rangle
-\mathrm{e}^{\mathrm{i}\theta(s)} |\!\!\uparrow\downarrow\rangle +\sqrt{2}|\!\!\downarrow \uparrow \rangle\!\right)\!\!,
\label{newzeroonestate}\\
|1^0(s)\rangle\! &=&\! \frac{1}{2}\!\left(\mathrm{e}^{-\mathrm{i}\theta(t)}|\!\!\uparrow \uparrow \rangle
+|\!\!\uparrow\downarrow\rangle +\sqrt{2}|\!\!\downarrow\downarrow\rangle\!\right)\!\!,
\label{newonezerostate}\\
|1^1(s)\rangle\! &=&\! \frac{1}{2}\!\left(\!|\!\! \uparrow\uparrow \rangle 
-\mathrm{e}^{\mathrm{i}\theta(s)}|\!\!\uparrow\downarrow\rangle -\sqrt{2}|\!\!\downarrow \uparrow
\rangle\!\right)\!\!. \label{newoneonestate}
\end{eqnarray} 

An arbitrary state in the standard basis
\begin{equation}
|\Psi(s)\rangle = \sum_{i,j=\downarrow,\uparrow}c_{ij}(s)|ij\rangle
\label{state1}
\end{equation}
when inserted into SE (\ref{SET}) leads to the following set of coupled
differential equations,
\begin{eqnarray}
\mathrm{i}\epsilon\dot{c}_{\uparrow\uparrow}(s) &=& -c_{\downarrow\uparrow}(s) 
+\mathrm{e}^{-\mathrm{i}\theta(s)}c_{\downarrow\downarrow}(s),\label{dif1}\\
\mathrm{i}\epsilon\dot{c}_{\uparrow\downarrow}(s) &=&\mathrm{e}^{\mathrm{i}\theta(s)}c_{\downarrow\uparrow}(s) 
+c_{\downarrow\downarrow}(s),\\
\mathrm{i}\epsilon\dot{c}_{\downarrow\uparrow}(s) &=& -c_{\uparrow\uparrow}(s) 
+\mathrm{e}^{-\mathrm{i}\theta(s)}c_{\uparrow\downarrow}(s),\\
\mathrm{i}\epsilon\dot{c}_{\downarrow\downarrow}(s) &=&\mathrm{e}^{\mathrm{i}\theta(s)}c_{\uparrow\uparrow}(s) 
+c_{\uparrow\downarrow}(s),\label{dif4}
\end{eqnarray}
where
\begin{equation}
\epsilon(s) = \sqrt{2}\hbar v/E(s).
\label{eps}
\end{equation}

We assume that the system starts at the ground state $|0^0(0)\rangle$ which 
gives the following initial conditions  
$c_{\uparrow\uparrow}(0) =\mathrm{e}^{-\mathrm{i}\theta_0}/2$, $c_{\uparrow\downarrow}(0) = 1/2$,
$c_{\downarrow\uparrow}(0) = 0$, and $c_{\downarrow\downarrow}(0) = -\sqrt{2}/2$. 

To compare the exact time-evolved state with the corrections
coming from DAPT it is better to express Eq.~(\ref{state1}) 
in terms of the snapshot eigenvectors 
(\ref{newzerozerostate})-(\ref{newoneonestate}),
\begin{equation}
|\Psi(s)\rangle = \sum_{i,j=0,1}d_{ij}(s)|i^j(s)\rangle,
\label{state2}
\end{equation}
where
\begin{eqnarray}
&&\hspace*{-0.5cm}d_{00}(s) = (\mathrm{e}^{\mathrm{i}\theta(s)}c_{\uparrow\uparrow}(s)+c_{\uparrow\downarrow}(s)
-\sqrt{2}c_{\downarrow\downarrow}(s))/2,\\
&&\hspace*{-0.5cm}d_{01}(s)= (c_{\uparrow\uparrow}(s)-\mathrm{e}^{-\mathrm{i}\theta(s)}c_{\uparrow\downarrow}(s)
+\sqrt{2}c_{\downarrow\uparrow}(s))/2,\\
&&\hspace*{-0.5cm}d_{10}(s) = (\mathrm{e}^{\mathrm{i}\theta(s)}c_{\uparrow\uparrow}(s)+c_{\uparrow\downarrow}(s)
+\sqrt{2}c_{\downarrow\downarrow}(s))/2,\\
&&\hspace*{-0.5cm}d_{11}(s) = (c_{\uparrow\uparrow}(s)-\mathrm{e}^{-\mathrm{i}\theta(s)}c_{\uparrow\downarrow}(s)
-\sqrt{2}c_{\downarrow\uparrow}(s))/2.
\end{eqnarray}

We measure the closeness between the exact solution (\ref{state2}), numerically obtained
by solving Eqs.~(\ref{dif1})-(\ref{dif4}), and the states $|\Psi(s)\rangle_{N_k}$,
via the infidelity \cite{Rig08}
\begin{equation}
I_k(s) = 1 - |\langle\Psi(s)|\Psi(s)\rangle_{N_k}|^2.
\label{inf}
\end{equation}
Here $|\Psi(s)\rangle_{N_k}$ is the normalized state with terms up to
order $k$ obtained from DAPT,
$$
|\Psi(s)\rangle_{N_k} = N_k(s) \sum_{p=0}^{k}v^p|\Psi^{(p)}(s)\rangle,
$$
with $N_k(s)$ being a normalization factor, and $0\leq I_k(s)\leq1$. The smaller
$I_k(s)$ the closer $|\Psi(s)\rangle_{N_k}$ is to the exact solution
while for $I_k(s)= 1$ they become orthogonal.    

The state $|\Psi^{(p)}(s)\rangle$ is obtained solving the recursive 
relation (\ref{recursiveB}) with the initial conditions (\ref{initialvector}),
(\ref{B0}), and (\ref{Bnn0}).
After that, we pick the first element of the vector (\ref{ansatzA}) leading 
to the state $|\Psi^{(p)}(s)\rangle$ as given above.

The first case we study is the one with $\lambda=0$, i.e., the case with a constant
gap. In contrast to the exactly solvable model of Sec. \ref{example}, 
where we also had a constant gap, now the time
dependence of the Hamiltonian is quadratic in $s$. 

Building on previous knowledge and similar examples for
non-degenerate systems \cite{Rig08} we expect that the quality of DAPT will 
depend on the interplay between the parameter
$v$ and the minimum gap between the ground and excited eigenspaces; the smaller the gap
the smaller $v$ must be for DAPT to provide meaningful results.  Therefore, looking at Eq.~(\ref{eps})
we realize that whenever $\epsilon(s)\ll 1$ DAPT is supposed to give accurate results.

\begin{figure}[!ht]
\includegraphics[width=8cm]{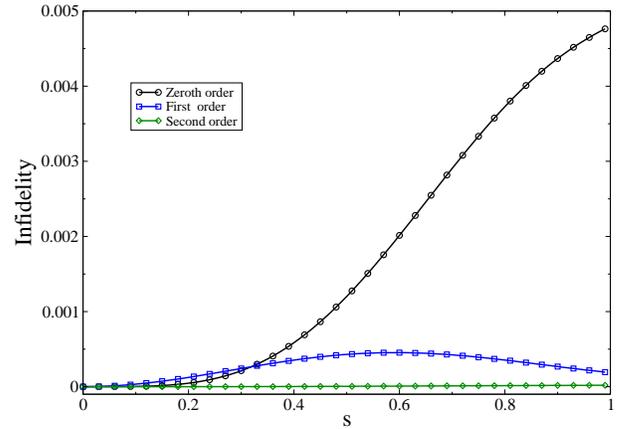}
\caption{\label{fig2} (Color online) Here we work with a constant gap ($\lambda = 0$) and
$\hbar = 1.0$, $\theta_0=0.1$, $E_0=1.5$ and $v=w=0.5$,
giving $\epsilon \approx 0.47$. Note that by including higher order terms in the perturbative
series we get a better description of the evolved state. Here and in the following 
figures all quantities are dimensionless.}
\end{figure}

\begin{figure}[!ht]
\includegraphics[width=8cm]{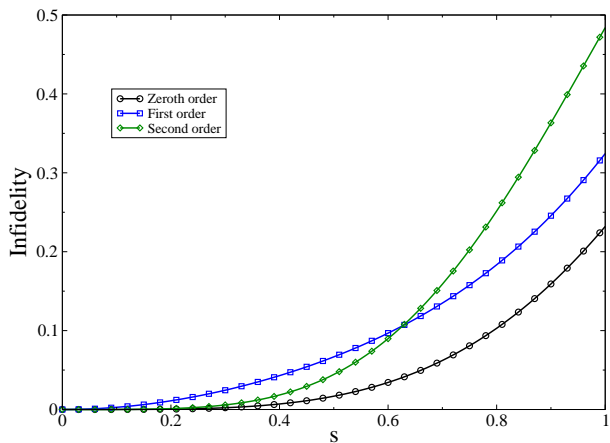}
\caption{\label{fig3} (Color online) Here $\lambda=0$, $\hbar = 1.0$, $\theta_0=0.1$, $E_0=1.5$ and $v=w=1.5$
giving $\epsilon \approx 1.41$. Now the inclusion of higher order terms is 
detrimental since $\epsilon>1.$}
\end{figure}

This is indeed the case as Figs. \ref{fig2} and \ref{fig3} illustrate.
For $\epsilon \approx 0.5$ (Fig. \ref{fig2}) we notice that the more
orders we include in the perturbation series the better. Also, by 
just going up to second order in $v$ we already get an excellent description of the
exact solution. On the other hand, for $\epsilon \approx 1.4$ (Fig. \ref{fig3}) 
we observe, as expected, the break down of DAPT.    

\begin{figure}[!ht]
\includegraphics[width=8cm]{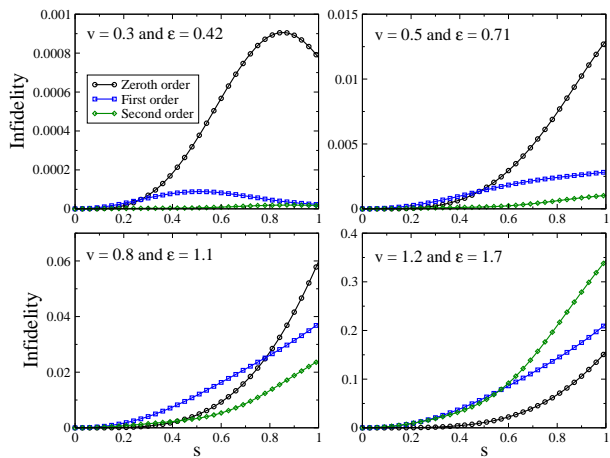}
\caption{\label{fig4} (Color online) Here $\lambda=\hbar = 1.0$, $\theta_0=0.1$, $E_0=1.0$ and $v=w$.
Note that as $\epsilon \rightarrow 1$ the perturbative series breaks down.}
\end{figure}

\begin{figure}[!ht]
\includegraphics[width=8cm]{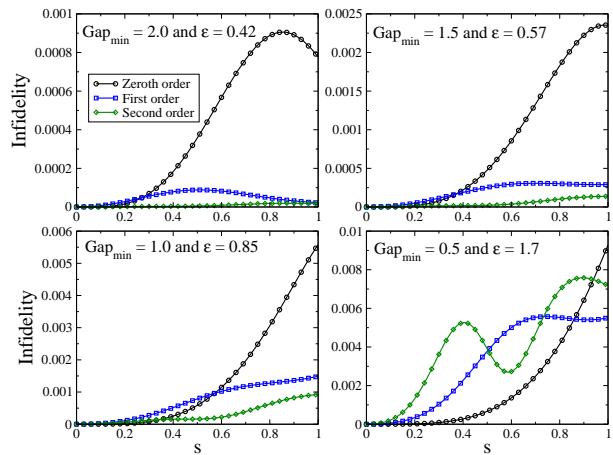}
\caption{\label{fig5} (Color online) Here $\lambda=\hbar = 1.0$, $\theta_0=0.1$, and $v=w=0.3$.
Note that as the gap ($2E_0$) increases the better DAPT describes the evolved state.
For $\epsilon \gtrsim 1$ it no longer works. In the figure we see the values
of $\epsilon$ calculated for the minimum gap.
}
\end{figure}

Let us now work with a time-dependent gap, which can be 
achieved by setting $\lambda=1$ in Eq.~(\ref{lamb}). 
Within this choice of $\lambda$, we deal with two different scenarios. 
First we fix the  minimum gap ($2E_0$) and successively solve 
Hamiltonian (\ref{otherH}) for increasing $v$ (Fig. \ref{fig4}). 
Next we fix $v$ and solve (\ref{otherH}) for different values of
$E_0$ (Fig. \ref{fig5}).   

In Fig. \ref{fig4} we see that for all values of $v$ such that $\epsilon < 1$ 
(upper panels), the more orders we include in the perturbative series the closer we get
to the exact solution. And the lower $v$ the better the approximation. 
By increasing $v$ we arrive at a point where $\epsilon > 1$ (lower panels) and
we start to see the breaks down of DAPT, which becomes more manifest for greater values of $\epsilon$. 

Finally, in Fig. \ref{fig5} we clearly note that, for fixed $v$, the greater the minimum gap
the better DAPT. By continually decreasing the gap we keep increasing $\epsilon$ until it
gets larger than one. In such a case, as can be seen in the lower-right panel of Fig. \ref{fig5},
DAPT breaks down. For values of $\epsilon\approx 1$, but still lower than one, we need to
include higher orders to get a good approximation; just keeping terms up to
first order is not enough to outperform the zeroth order approximation
during the whole time evolution (lower-left panel of Fig. \ref{fig5}).        

\section{Conclusions}
\label{conclusion}

We presented and expanded on the degenerate adiabatic perturbation theory (DAPT) first
introduced in Ref. \cite{Rig10}, whose goal is to provide consistent perturbative corrections about the
degenerate adiabatic approximation (DAA) for time-dependent systems.
We provided all the missing mathematical steps leading 
to its development as well as new physical insights and a better understanding of concepts
brought forth by DAPT. In particular, we emphasized the importance of 
three key ingredients without which the development of DAPT would be doomed to failure.

First, we showed the importance of a proper rescaling of the time $s=vt$ in the Schr\"odinger
equation (SE) in terms of the adiabatic parameter $v=1/T$, which is related to the rate 
(``velocity'') at which the Hamiltonian $\mathbf{H}(t)$ is driven from its initial to its 
final configuration, and upon which the perturbative series is built. 
On the formal level, this rescaling allowed us to properly identify the correct order
 in the perturbative series. 

This  rate depends on intrinsic internal parameters of $\mathbf{H}(t)$
which ultimately reflects on how $\mathbf{H}(t)$ changes along the parameter space in
order to reach a given final configuration. 
Indeed, it is a delicate balance between the rate at which $\mathbf{H}(t)$
changes and the duration $T$ of the whole time evolution (experiment) what
dictates whether DAPT converges or not and, hence, furnishes meaningful perturbative 
corrections about DAA. Reversing the argument of the previous sentence, a 
failure of DAPT to converge may indicate that for a given $\mathbf{H}(t)$
and experimental running time $T$ the system's evolution cannot be approximated 
by DAA. In such a case, for the system to be well approximated by DAA and the leading orders of DAPT,
we should slow down the rate at which $\mathbf{H}(t)$ changes by decreasing $v$, which manifests itself
in a longer experimental running time (greater $T$).

Second, in order to make any progress we had to propose the right ansatz with
a compact and clear notation for ease of later mathematical manipulations; one that
had at the same time the following two characteristics. On one hand it should 
correctly deal with the rescaled time $s$ and perturbative
parameter $v$. As such, it should factor out terms of order $\mathcal{O}(v^{-1})$ 
and below, which are the problematic ones when $v\rightarrow 0$, i.e., when
the system's evolution should be well described by DAA. On the
other hand, the ansatz should lead to computable higher order corrections in a 
straightforward and numerically robust manner. The ansatz we presented
had these two properties allowing us to get recursive relations where the
$p$-th order correction (the term multiplying $v^p$ in the perturbative series, $p\geq 1$) 
is obtained from knowledge of the order $p-1$.

Third, in contrast to many standard time-dependent perturbation theories,
the right ansatz alone is not enough to guarantee 
 the right perturbative corrections. It is of paramount importance to set the correct initial condition in the recursive
relations. This is accomplished by imposing that at $t=0$ all higher order
terms in the perturbative series expansion are zero with the exception of
the zeroth order, which is tuned to satisfy the system's initial condition.
Also, for the rest of the system's time evolution we must have that the zeroth order 
is DAA. This is achieved by building the zeroth order term in such a way that
transitions among different eigenspaces are forbidden. With this choice 
the non-abelian Wilczek-Zee (WZ) geometric phase naturally appears
as the solution to the zeroth order recursive relation.

In addition to the formal development of DAPT, which allowed us to
correctly determine perturbative corrections about DAA, the ideas
summarized in the previous paragraphs paved the ground to a rigorous formulation of
the adiabatic theorem of quantum mechanics for non-degenerate and
degenerate systems. With the aid of DAPT, we were able 
to formulate and prove necessary and sufficient conditions for the validity of
the degenerate adiabatic theorem (DAT) \cite{Rig12}. A more extensive
discussion of DAT and its physical meaning, in particular the notion of 
\textit{slowly} changing Hamiltonians, as well as all 
technical details of the proofs outlined in \cite{Rig12} were
presented in Sec. \ref{NSC}. 

In the remaining sections of this paper we applied both DAPT and 
the conditions for the validity of DAT to a few examples. We
derived in full detail   the exact closed form solution \cite{Rig10} to 
a degenerate time-dependent problem \cite{Bis89}, which is a natural 
extension of the famous non-degenerate spin-1/2 system subjected to a rotating external 
magnetic field \cite{Rab54}. We then verified that DAPT gives the correct 
perturbative corrections to this model, matching exactly the expansion of the exact
solution in terms of $v$. 
Furthermore, we applied the necessary and sufficient conditions for the validity of DAT
to this model and showed that they give the correct 
conditions under which DAA is a good description for the system's evolution. 
We then solved numerically several time-dependent
Hamiltonians in order to compare their solutions 
with the perturbative series derived from DAPT. We showed that for small
enough $v$ DAPT gives excellent results by just 
truncating the perturbative series at the second order. Finally, we should mention that
the study of both the exactly solvable model and the numerical ones allowed us
to have a better grasp of the meaning of $v$ and also understand  the conditions under which 
DAPT provides meaningful results.

\begin{acknowledgments}
GR thanks for funding the Brazilian agencies CNPq 
(National Council for Scientific and Technological Development), 
FAPESP (State of S\~ao Paulo Research Foundation), and INCT-IQ (National Institute of
Science and Technology for Quantum Information).
\end{acknowledgments}

\appendix

\section{The Wilczek-Zee Phase}
\label{appendixA}

We can write the most
general state describing a degenerate system as (cf. Eq. (\ref{vector0})),
\begin{equation}
\mathbf{|\Psi}(t)\rangle = \sum_{n=0}
\mathrm{e}^{-\mathrm{i}\omega_n(t)}
b_n(t)\mathbf{U}^{n}(t)\mathbf{|n}(t)\rangle, 
\label{ap1}
\end{equation}
with the dynamical phase $\omega_n(t)$ given by Eq. (\ref{dynamical})
(for the moment we make no assumptions about the other quantities).
To proceed, we need to work with the transposed quantities
(cf. Sec. \ref{DAPT}). Thus, 
\begin{equation}
\mathbf{|\Psi}(t)\rangle^T = \sum_{n=0}
\mathrm{e}^{-\mathrm{i}\omega_n(t)}
b_n(t)\mathbf{|n}(t)\rangle^T\mathbf{U}^{n}(t)^T. 
\label{ap2}
\end{equation}

Transposing SE, Eq. (\ref{SE}), inserting Eq.~(\ref{ap2}) into
it, and left multiplying both sides by $\langle\mathbf{m}(t)|^T$,
we get after exchanging $n\leftrightarrow m$ and transposing back,
\begin{eqnarray}
\dot{b}_n(t)\mathbf{U}^{n}(t) & + & b_n(t)\dot{\mathbf{U}}^n(t)
\nonumber \\
& + & \sum_{m=0}b_m(t)\mathrm{e}^{-\mathrm{i}\omega_{mn}(t)}
\mathbf{U}^{m}(t)
\mathbf{M}^{mn}(t)=0,\nonumber \\
\label{ap3}
\end{eqnarray}
with $\omega_{mn}(t)=\omega_m(t) - \omega_n(t)$ and
$\mathbf{M}^{mn}(t)$ given by Eq.~(\ref{M}).
So far no approximation has been made and, in principle, the time
evolution could be determined by solving the system of  differential
equations given in (\ref{ap3}).

The degenerate adiabatic approximation (DAA) consists in neglecting the coupling
between \textit{different} eigenspaces $\mathcal{H}_n$ but not those
within a given eigenspace, i.e., we must have
\begin{equation}
\mathbf{M}^{nm}(t) \approx \delta_{nm}\mathbf{M}^{nn}(t) 
\hspace{.5cm}\text{and}\hspace{.5cm} 
b_n(t) \approx b_n(0).
\label{ap4}
\end{equation}
Inserting Eq.~(\ref{ap4}) into (\ref{ap3}) gives,
\begin{equation}
\dot{\mathbf{U}}^n(t) - \mathbf{U}^{n}(t)\mathbf{A}^{nn}(t)=0,
\label{difWZ}
\end{equation}
where we defined $\mathbf{A}^{nm}(t)=-\mathbf{M}^{nm}(t)$. The previous
differential equation is nothing but the Wilczek-Zee (WZ) phase, whose
formal solution is written as \cite{Wil84}
\begin{equation}
\mathbf{U}^{n}(t) = \mathbf{U}^{n}(0)\mathcal{T}
\exp\left( \int_0^t\mathbf{A}^{nn}(t')dt'\right),
\end{equation}
with $\mathcal{T}$ denoting time-ordering. We should note that
in Ref. \cite{Wil84} the authors assume that the system is 
initialized in an  eigenvector of $\mathbf{H}(t)$. 
Here we relax this assumption (Eq.~(\ref{ap1})). It is by using conditions  
(\ref{ap4}) that we derive the WZ phase and establish that
the system evolves according to Eq.~(\ref{vector0}).

The previous approach, however, does not provide a rigorous way to
get necessary and sufficient conditions guaranteeing 
that the system's evolution can be approximated by Eq.~(\ref{vector0}).
And the reason is simple: in general we do not have the solution
to SE that leads to the explicit formula for $b_n(t)$. 
All we can do is to test whether
the first piece of  Eq.~(\ref{ap4}), 
the one that  can be computed without knowing the solution
to the SE, is a valid approximation. In other words, if for $n\neq m$ and all $t$
$$
\mathbf{M}^{nm}(t)\ll 1 ,
$$
then DAA is probably a good
description of the system's evolution. But we can get into trouble
because this does not necessarily imply 
$$
\dot{b}_n(t)\ll 1.
$$
Rigorous conditions
and how much we are losing by neglecting higher order terms
can be obtained, though, by using DAPT (cf. Secs. \ref{DAPT} and \ref{NSC}).
                              
\section{Proof that DAPT implies APT}
\label{appendixB}

The non-degenerate ansatz of APT as given in \cite{Rig08} is
\begin{equation}
|\Psi(s)\rangle = \sum_{n,m=0}\sum_{p=0}^{\infty}
v^p \mathrm{e}^{-\frac{\mathrm{i}}{v}\omega_{m}(s)}
\mathrm{e}^{\mathrm{i}\gamma_{m}(s)}
b_{nm}^{(p)}(s)|n(s)\rangle,
\label{nansatz2}
\end{equation}
with $\gamma_m(s)$ the Berry phase, i.e., Eq.~(\ref{WZphase})
when no degeneracy is present. In the present notation, 
this ansatz led to the following recursive relation \cite{Rig08}
\begin{eqnarray}
\frac{\mathrm{i}}{\hbar}\Delta_{nm}(s)b_{nm}^{(p+1)}(s) &+& \dot{b}^{(p)}_{nm}(s)
-[\mathbf{M}^{mm}(s)]_{00}b_{nm}^{(p)}(s) \nonumber \\
&+& \sum_{k=0}b_{km}^{(p)}(s)[\mathbf{M}^{kn}(s)]_{00}=0,
\label{nrecursive1}
\end{eqnarray}
with the following zeroth order term
\begin{equation}
b_{nm}^{(0)}(s) = b_n(0)\delta_{nm}. 
\label{nbn0s}
\end{equation}

To prove that APT is a particular case of DAPT we need to show that 
Eqs.~(\ref{nansatz2})-(\ref{nbn0s}) are equivalent to the ones derived from
DAPT when we assume no-degeneracy.

First, comparing Eq.~(\ref{nansatz2}) with DAPT ansatz (\ref{ansatz2})
we note that we must have
\begin{equation}
\mathbf{B}_{mn}^{(p)}(s)=\mathrm{e}^{\mathrm{i}\gamma_{m}(s)}
b_{nm}^{(p)}(s)
\label{Bb}
\end{equation}
for them to be equivalent. 
Inserting Eq.~(\ref{Bb}) into (\ref{nrecursive1}) and noting that
$\dot{\gamma}_m(s)=\mathrm{i}[\mathbf{M}^{mm}(s)]_{00}$ leads to
\begin{equation}
\frac{\mathrm{i}}{\hbar}\Delta_{nm}(s)\mathbf{B}_{mn}^{(p+1)}(s) 
+ \dot{\mathbf{B}}^{(p)}_{mn}(s) \nonumber \\
+ \sum_{k=0}\mathbf{B}_{mk}^{(p)}(s)\mathbf{M}^{kn}(s)=0,
\end{equation}
which is exactly the recursive relation (\ref{recursiveB}) of DAPT.

Finally, for $p=0$ if we insert (\ref{nbn0s}) into (\ref{Bb}) 
we get DAPT zeroth order noting that for non-degenerate systems 
$\mathbf{U}^{n}(s)=\mathrm{e}^{\mathrm{i}\gamma_n(s)}$.

\section{General solution to the recursive relation}
\label{appendixC}

Our goal here is to manipulate Eq.~(\ref{recursiveB})  to 
explicitly obtain $\mathbf{B}_{mn}^{(p+1)}(s)$ in terms of the lower order
coefficients. For $n\neq m$,
Eq.~(\ref{recursiveB}) straightforwardly implies
\begin{eqnarray}
\mathbf{B}_{mn}^{(p+1)}(s)
&=& \frac{\mathrm{i}\hbar}{\Delta_{nm}(s)}\left(
\dot{\mathbf{B}}_{mn}^{(p)}(s)
+\sum_{k=0}\mathbf{B}_{mk}^{(p)}(s)
\mathbf{M}^{kn}(s)\right).
\nonumber \\ 
\label{C1}
\end{eqnarray}

When $n=m$,  
Eq.~(\ref{recursiveB}) gives
\begin{eqnarray}
\dot{\mathbf{B}}_{nn}^{(p+1)}(s)
&+&\mathbf{B}_{nn}^{(p+1)}(s)\mathbf{M}^{nn}(s)\nonumber \\
&+&\mathop{\sum_{k=0}}_{k\neq n}
\mathbf{B}_{nk}^{(p+1)}(s)\mathbf{M}^{kn}(s)
=0.
\end{eqnarray}
Making the following change of variable
\begin{equation}
\mathbf{B}_{nn}^{(p+1)}(s) =
\mathbf{\tilde{B}}_{nn}^{(p+1)}(s)\mathbf{U}^n(s)
\end{equation}
leads to
\begin{eqnarray}
&&\mathbf{\tilde{B}}_{nn}^{(p+1)}(s)\!
\left(\! \dot{\mathbf{U}}^n(s) +
\mathbf{U}^n(s)\mathbf{M}^{nn}(s)
\!\right)
+
\mathbf{\dot{\tilde{B}}}_{nn}^{(p+1)}(s)
\mathbf{U}^n(s)
\nonumber \\
&+&\mathop{\sum_{k=0}}_{k \neq n}
\mathbf{B}_{nk}^{(p+1)}(s)
\mathbf{M}^{kn}(s)
= 0.
\end{eqnarray}
The term inside the parenthesis is zero since 
$\mathbf{U}^n(s)$ is the WZ-phase (cf. Eq.~(\ref{difWZ})). 
Then, using the unitarity of $\mathbf{U}^n(s)$, we can solve
for $\mathbf{\tilde{B}}_{nn}^{(p+1)}(s)$,
\begin{widetext}

\begin{equation}
\mathbf{\tilde{B}}_{nn}^{(p+1)}(s) = 
\mathbf{\tilde{B}}_{nn}^{(p+1)}(0)
-\mathop{\sum_{m=0}}_{m \neq n}\int_0^s
\mathbf{B}_{nm}^{(p+1)}(s')
\mathbf{M}^{mn}(s')
\left(\mathbf{U}^n(s')\right)^\dagger
\mathrm{d}s',
\end{equation}
where we have changed $k \rightarrow m$.
Then, returning to the original variable 
\begin{eqnarray}
\mathbf{B}_{nn}^{(p+1)}(s) = 
-\mathop{\sum_{m=0}}_{m \neq n} 
\mathbf{B}_{mn}^{(p+1)}(0)
\left(\mathbf{U}^n(0)\right)^\dagger
\mathbf{U}^n(s)
- \mathop{\sum_{m=0}}_{m \neq n}
\int_0^s\mathrm{d}s'
\left(
\mathbf{B}_{nm}^{(p+1)}(s')
\mathbf{M}^{mn}(s')
\left(\mathbf{U}^n(s')\right)^\dagger
\right)
\mathbf{U}^n(s),
\end{eqnarray}
where we have written the initial condition (\ref{B0}) as
\begin{equation}
\mathbf{\tilde{B}}_{nn}^{(p+1)}(0) = -
\mathop{\sum_{m=0}}_{m \neq n}
\mathbf{\tilde{B}}_{mn}^{(p+1)}(0)
\left(\mathbf{U}^n(0)\right)^\dagger.
\end{equation}
Finally, using Eq.~(\ref{C1}) we get
\begin{eqnarray}
\mathbf{B}_{nn}^{(p+1)}(s) &=& - \mathrm{i}\hbar \mathop{\sum_{m=0}}_{m
\neq n}\frac{\dot{\mathbf{B}}_{mn}^{(p)}(0)
\left(\mathbf{U}^n(0)\right)^\dagger
\mathbf{U}^n(s)}{\Delta_{nm}(0)} 
-\mathrm{i}\hbar \mathop{\sum_{m=0}}_{m \neq n}\sum_{k=0}
\frac{\mathbf{B}_{mk}^{(p)}(0) \mathbf{M}^{kn}(0)
\left(\mathbf{U}^n(0)\right)^\dagger \mathbf{U}^n(s)}
{\Delta_{nm}(0)} \nonumber \\
&&+\mathrm{i}\hbar\mathop{\sum_{m=0}}_{m \neq n} \int_0^s \left( \frac{
\dot{\mathbf{B}}_{nm}^{(p)}(s') \mathbf{M}^{mn}(s')
\left(\mathbf{U}^n(s')\right)^\dagger}{\Delta_{nm}(s')} \right)
\mathrm{d}s' \mathbf{U}^n(s)
\nonumber \\
&&+\mathrm{i}\hbar\mathop{\sum_{m=0}}_{m \neq n} \sum_{k=0} \int_0^s
\left( \frac{\mathbf{B}_{nk}^{(p)}(s') \mathbf{M}^{km}(s')
\mathbf{M}^{mn}(s') \left(\mathbf{U}^n(s')\right)^\dagger}
{\Delta_{nm}(s')}\right) \mathrm{d}s' \mathbf{U}^n(s).
\label{C8}
\end{eqnarray}

Equations (\ref{C1}) and (\ref{C8}), together with the zeroth order term
(Eq. (\ref{B0})),
$
\mathbf{B}_{mn}^{(0)}(s) = b_n(0)\mathbf{U}^n(s)\delta_{mn},
$
are all that we need to get perturbative corrections about DAA to any order. However,
in many applications of DAPT, it is easier to use the recursive relation (\ref{recursiveB})
directly.

\end{widetext}

\section{$\theta \approx 0$ does not imply $|\Psi(t)\rangle \approx |\Psi^{(0)}(t)\rangle$}
\label{appendixD}

Expanding up to first order in $\theta$ DAA, Eq. (\ref{exampleAd}), and the 
exact solution, Eq. (\ref{ground_exact}), we get
respectively
\begin{eqnarray*}
|\Psi^{(0)}(t)\rangle &=& \mathrm{e}^{\frac{\mathrm{i} b t}{2}}|0^0(t)\rangle + 
\mathrm{i} \theta \mathrm{e}^{-\frac{1}{2} \mathrm{i} (w-b) t} \sin \left(\frac{w t}{2}\right)|0^1(t)\rangle, \\
|\Psi(t)\rangle &=& |\Psi^{(0)}(t)\rangle + \frac{\mathrm{i} \theta w \mathrm{e}^{-\frac{1}{2} \mathrm{i} w t } 
\sin \left(\frac{1}{2} (w-b) t\right)}{w-b} |1^1(t)\rangle.
\end{eqnarray*}

Comparing both expressions it is clear that if $w\geq b$ the probabilities to measure $|0^1(t)\rangle$ and
$|1^1(t)\rangle$ are always of the same order in $\theta$ and, therefore, 
the system cannot be properly described by DAA.
However, when $w\ll b$ it is clear that the probability to get $|1^1(t)\rangle$ vanishes and the one to
obtain $|0^1(t)\rangle$ does not (there is no $b$ in its denominator). This shows, as expected, 
that for slowly rotating fields DAA is a good approximation to the system's evolution.


\begin{thebibliography}{99}

\bibitem{Coh77} C. Cohen-Tannoudji, B. Diu, and F. Lalo\"e, \textit{Quantum Mechanics}
(John Wiley \& Sons, New York, 1977), vol. 2.

\bibitem{Mes62} A. Messiah, \textit{Quantum Mechanics}
(North-Holland, Amsterdam, 1962), vol. 2.

\bibitem{Jan07} S. Jansen, M.-B. Ruskai, and R. Seiler, J. Math. Phys. \textbf{48}, 102111 (2007).

\bibitem{Rig08} G. Rigolin, G. Ortiz, and V. H. Ponce,
Phys. Rev. A \textbf{78}, 052508 (2008).

\bibitem{Pol10a} C. De Grandi, V. Gritsev, and A. Polkovnikov,  Phys. Rev. B \textbf{81}, 012303 (2010).

\bibitem{Pol10b} C. De Grandi and A. Polkovnikov, Lecture Notes in Physics \textbf{802}, 75 (2010).

\bibitem{Pol11a} C. De Grandi, A. Polkovnikov, and A. W. Sandvik, Phys. Rev. B \textbf{84}, 224303 (2011).

\bibitem{Pol11b} A. Polkovnikov, K. Sengupta, A. Silva, and M. Vengalattore, Rev. Mod. Phys. \textbf{83}, 863 (2011).  

\bibitem{Rig10} G. Rigolin and G. Ortiz, Phys. Rev. Lett. \textbf{104},
170406 (2010).

\bibitem{Ton10} D. M. Tong,  Phys. Rev. Lett.  \textbf{104}, 120401 (2010).

\bibitem{Rig12} G. Rigolin and G. Ortiz, Phys. Rev. A \textbf{85}, 062111 (2012).

\bibitem{Iva01}  D. A. Ivanov, Phys. Rev. Lett.  \textbf{86}, 268 (2001).

\bibitem{Cob14} E. Cobanera and G. Ortiz, Phys. Rev. A \textbf{89}, 012328 (2014).

\bibitem{Una98} R. Unanyan, M. Fleischhauer, B. W. Shore, and K. Bergmann, Optics Commun. \textbf{155}, 144 (1998).

\bibitem{Una99} R. G. Unanyan, B. W. Shore, and K. Bergmann, Phys. Rev. A \textbf{59}, 2910 (1999).

\bibitem{Kis04} Z. Kis, A. Karpati, B. W. Shore, and N. V. Vitanov, Phys. Rev. A \textbf{70}, 053405 (2004).

\bibitem{Tha04} I. Thanopulos, P. Kr\'al, and M. Shapiro, Phys. Rev. Lett. \textbf{92}, 113003 (2004).

\bibitem{Bis89} S. N. Biswas, Phys. Lett. B \textbf{228}, 440
(1989).

\bibitem{Ber84} M. V. Berry, Proc. R. Soc. Lond. A \textbf{392}, 45 (1984).

\bibitem{Wil84} F. Wilczek and A. Zee, Phys. Rev. Lett. \textbf{52},
2111 (1984).

\bibitem{Wil11} F. Wilczek, Opening talk at Nobel Symposium 148, 
eprint: arXiv:1109.1523v1 [cond-mat.mes-hall].

\bibitem{Tho83} D. J. Thouless, Phys. Rev. B \textbf{27}, 6083 (1983).

\bibitem{Che11} D. Cheung, P. H{\o}yer and N. Wiebe, J. Phys. A: Math. Theor. \textbf{44}, 415302 (2011). 

\bibitem{footnote1} Due to the Kronecker delta appearing in
Eq.~(\ref{B0}) the rhs of (\ref{newsuf}) can also be written as
$\sum_{m=0}|[\mathbf{B}_{mn}^{(0)}(s)]_{0g_n}|$. This is how
it is presented in \cite{Rig12}.


\bibitem{Rab54} I. I. Rabi, N. F. Ramsey, and J. Schwinger, Rev. Mod. Phys.
\textbf{26}, 167 (1954).

\bibitem{Boh93} A. Bohm, \textit{Quantum Mechanics: Foundations and
Applications} (Springer-Verlag, New York, 1993), p. 587.

\bibitem{Yuk09} V. I. Yukalov, Phys. Rev. A \textbf{79}, 052117 (2009).

\end{thebibliography}
\end{document}